\documentclass[twocolumn]{aastex631}

\shorttitle{Transit Depth Variations in TOI-216}
\shortauthors{McKee and Montet}
\graphicspath{{./}{figures/}}

\begin{document}

\title{Transit Depth Variations Reveal TOI-216\,b to be a Super-Puff}

\correspondingauthor{Brendan J. McKee}
\email{b.mckee@unsw.edu.au}

\author[0000-0001-8421-2833]{Brendan~J.~McKee}
\affiliation{School of Physics, University of New South Wales, Sydney, NSW 2052, Australia}

\author[0000-0001-7516-8308]{Benjamin~T.~Montet}
\affiliation{School of Physics, University of New South Wales, Sydney, NSW 2052, Australia}
\affiliation{UNSW Data Science Hub, University of New South Wales, Sydney, NSW 2052, Australia}



\begin{abstract}

The planets of the TOI-216 system have been previously observed to exhibit large transit timing variations (TTVs), which enabled precise mass characterization of both transiting planets. In the first year of \textit{TESS} observations, TOI-216\,b exhibited grazing transits, precluding a measurement of its radius. In new observations, we demonstrate the orbit of the planet has precessed and it is now fully transiting, so we can accurately measure its radius. TOI-216\,b is a puffy Neptune-mass planet, with a much larger radius that is now well constrained to $7.84^{+0.21}_{-0.19}$ $R_{\earth}$ and a density of $0.201\pm0.017$ g cm$^{-3}$. We numerically integrate the system across the \textit{TESS} observations to update and refine the masses and orbits of both planets, finding the uncertainty in the masses are now dominated by uncertainties in the stellar parameters. TOI-216\,b represents a growing class of super-puff planets in orbital resonances and with a companion in a nearly circular orbit, suggesting the early evolution of these planets is driven by smooth disk migration.


\end{abstract}



\section{Introduction} \label{sec:intro}

Warm Jupiters are a class of gas giant exoplanets with 10-100 day orbital periods. These planets orbit further from their host star than hot Jupiters and are expected to have somewhat differing migration methods to explain their location \citep{Dawson_2015}. The two main migration paths for hot Jupiters are thought to be disk migration \citep{Goldreich_1980, Lin_1986} and high-eccentricity tidal migration \citep{Fabrycky_2007}. In warm Jupiter systems, \citet{Huang_2016} argue that high-eccentricity migration is less common, and in situ formation becomes an alternative method of placing these large planets at the resulting distance. Characterizing the radii and masses of such planets provides information about their bulk composition \citep{Chabrier_2009, Lopez_2014}, and dynamical modeling allows for the orbital properties to be determined \citep{Agol_2018}. 

The most prolific method of detecting exoplanets has been the transit method \citep{Deeg_2018}. Measuring the transit depth allows for the radius of the planet to be determined, but offers no direct information about the mass of the planet \citep{Winn_2010}. In multi-planet systems, the gravitational pull of the planets on each other perturb the planets away from a simple Keplerian orbit. This can cause transit timing variations (TTVs) to be measured as the transit arrives earlier or later than expected for a strictly periodic signal \citep{Agol_2005}, such as observed in Kepler-9 \citep{Holman_2010} and Kepler-11 \citep{Lissauer_2011}. In systems where the ratios of planet periods are near to a mean motion resonance the transits can be observed many hours away from the expected transit time for a constant period \citep{Nesvorny_2013, Hamann_2019}. The amplitudes of the TTVs of each planet depend on the ratio of the masses of the planets as they perturb each other \citep{Nesvorny_2016}. The presence of TTVs in the light curve of a particular planet allows for constraints on the masses of other planets in the system \citep{Hadden_2017}. In a system with multiple transiting planets, the TTVs observed in each planet enable inference of constraints on the masses of each other planet. If both the mass and radius of a planet are known then the density can be calculated and the bulk composition determined \citep{Weiss_2014}.

For systems that are near mean motion resonance, the variation in transit timing changes sinusoidally in comparison to the expected transit time according to a super-period as calculated by the periods of each planet \citep{Lithwick_2012}. If the resonance is closer to a small integer ratio, successive conjunctions differ in mean longitude by a smaller amount, leading to a longer super-period and slower variations, as determined by Equation \ref{eq:super_period} for planets in a 2:1 orbital resonance. In this case, the super-period $P$ is given such that
\begin{equation}
P=\frac{1}{|2/P_c-1/P_b|}
\label{eq:super_period}
\end{equation}
where $P_b$ is the period of the inner planet and $P_c$ the period of the outer planet.

If the system is in a mean motion resonance, rather than near, the largest effect on TTVs will be due to the libration of the system around this resonance \citep{Deck_2015}. Libration occurs over a different period which scales with the mass of the planets \citep{Nesvorny_2016}. In order to best characterize the TTVs and the planets that create them, transits must be observed over a large enough portion of the dominant period such that the trend can be seen \citep{Hirano_2018, Hamann_2019}. If only a portion of the TTV cycle is observed the amplitude of the TTVs may not be apparent as the best fitting period to the available data may not be the best period over a longer timescale.

One system in which this is evident is TOI-216, which contains a transiting warm Jupiter and warm Neptune \citep{Dawson_2019, Kipping_2019}. This system is located close to the south ecliptic pole, allowing it to be observed for nearly all of the first and third years of the \textit{TESS} mission \citep{Ricker_2015}, as well as the second half of the fifth year. TOI-216 was initially characterized using TTVs measured from half a year of \textit{TESS} data, resulting in two solutions that differed in planetary masses, eccentricities and orbital resonances \citep{Dawson_2019}. This system was followed up when additional transits had been observed by \textit{TESS} at the start of the third year of the mission, along with ground-based transits and measurements of radial velocity. The additional data allowed one solution to be chosen, that of a warm Neptune, TOI-216\,b, in a 2:1 resonance with a warm Jupiter, TOI-216\,c \citep{Dawson_2021}. This solution precisely characterized the masses of each planet and radius of TOI-216\,c, but the measured radius of TOI-216\,b had a large uncertainty due to the grazing nature of the inner planet's transits. 

The \textit{TESS} mission has now observed TOI-216 across a four and a half year baseline, including two years of near continuous coverage. The addition of more transits to the dataset allows for the uncertainties on the characterization of the system to be narrowed. These new transits provide a longer TTV baseline to assist in determining masses, and carry information about transit depths that characterize radii. TOI-216\,b was previously noted to have grazing transits \citep{Dawson_2019, Kipping_2019}, where the depth of the transit is very sensitive to changes in impact parameter \citep{Miralda_Escude_2002}. Grazing transits are V-shaped and present a degeneracy between impact parameter and radius \citep{Gilbert_2022}. Fitting the change in impact parameters as seen in depth and duration variation adds another factor to confine the orbital elements of the system \citep{Masuda_2017, Mills_Fabrycky_2017, Dawson_2020}. This system was previously investigated for changes in impact parameter by \citet{Dawson_2020} using year 1 of \textit{TESS} data, where no significant change was detected. A later study by \citet{Dawson_2021} tentatively identified changes in the impact parameter of the inner planet using early year 3 \textit{TESS} data and ground based observations. We have seen in the most recent transits of TOI-216\,b that the transit depth has increased to a point where the planet is fully transiting, allowing for precise calculation of its radius. 

In this paper we use the \textit{TESS} light curve data to characterize the TOI-216 system, measuring the masses and radii of the two planets and inferring their densities. The increased observational baseline reveals that the inner planet is now fully transiting, confining its radius, with the change in impact parameter placing additional limits on the dynamics of the system. We achieve a precision in mass and radius that is limited by the uncertainty in stellar parameters. The calculated orbital geometry is used to make predictions about future transit times and changes in impact parameter that can be observed in future \textit{TESS} Sectors. The rest of this paper is organized as follows. In Section \ref{sec:data} we analyze and process the \textit{TESS} data to fit the transits in the light curves. In Section \ref{sec:analysis} we fit the transit timing variations and the variations in the impact parameter of the planets. We then examine the dynamics of the system in Section \ref{sec:results} to explore future changes in observable quantities. In Section \ref{sec:conclusions} we present our conclusions.

\section{Data} \label{sec:data}

TOI-216 was observed in \textit{TESS} Sectors 1 to 9 and 11 to 13 from 2018 July 25 to 2019 July 18 and Sectors 27 to 39 from 2020 July 4 to 2021 June 24, where it was positioned on Camera 4.  We used the two-minute cadence data as reduced by the Science Processing Operations Center pipeline \citep{Jenkins_2016}. We use the pre-search data conditioning simple aperture photometry light curves \citep{Smith_2012, Stumpe_2014}. A one-day long section around each transit was cut out from the data to separate the transits and allow them to be fit individually.

TOI-216 was also observed in \textit{TESS} Sectors 61 and 62, spanning 2023 January 18 to 2023 March 10.\footnote{\raggedright The target will continue to be observed throughout Cycle 5, appearing in every sector from 61 to 69, inclusive.} To produce a light curve for data from these sectors we downloaded the \textit{TESS} Image CAlibrator (TICA) Full-Frame Images as described by \citet{Fausnaugh_2020}, which represent observations of the field obtained at a 200-second cadence. We produced a light curve using a 3x3 pixel aperture centered on the target itself, with the least illuminated pixels from a larger 31x31 region centered on the target used to estimate the background flux levels, and we estimate the flux uncertainty from the scatter in the out-of-transit flux on transit timescales. In these data, two transits of planet b and one of transit c are visible and used in this analysis.

\begin{figure}[tb!]
\includegraphics[width=\columnwidth]{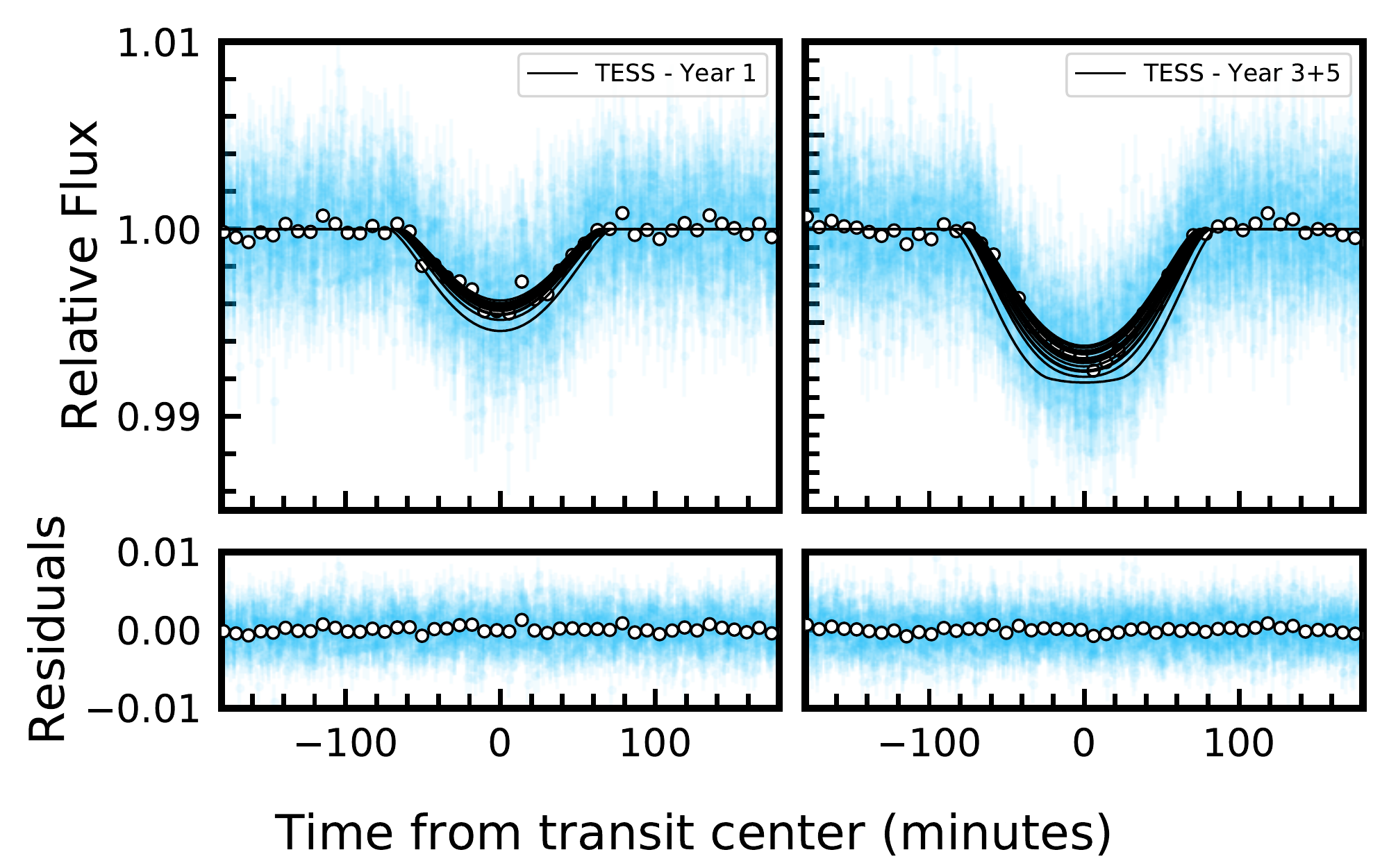}
\caption{Light curves of TOI-216\,b divided between year 1 (left) and year 3 onward (right)} of \textit{TESS} observations, phase aligned to the center of the transit. The best fitting model is shown in black for each individual transit. Binned data for each time interval is shown as white points.  
\label{fig:b_light_curves}
\end{figure}

We fit the light curves using the python \texttt{exoplanet} package \citep{exoplanet:joss}. Each transit was fit using common limb darkening coefficients $u_1$ and $u_2$ for both planets \citep{Kipping_2013}, while each planet was modeled using independent radii $R_b$ and $R_c$. The mass $M_\star$ and radius $R_\star$ of the star was allowed to vary following priors from \citet{Dawson_2021} inferred from stellar spectroscopy. In each transit the value of the impact parameter is allowed to vary, to investigate any changes in impact parameter over the observing period, as tentatively identified by \citet{Dawson_2021}. We model the impact parameters with a prior following \citet{Dawson_2020}, where they are drawn from a Cauchy distribution with mean $\overline{b}$ and change scale $\gamma$. To account for precession, we fix a different period, eccentricity, and argument of periapsis for generating the light curve model for transits in year 1, 3 and 5 based on the best fit to the transit times in each year. We sample each posterior distribution with \texttt{PyMC3} \citep{exoplanet:pymc3}, with 8,000 tuning steps and 40,000 draws. The reduced chi-squared value for the best fitting model is 0.95.

\begin{figure}[tb!]
\includegraphics[width=\columnwidth]{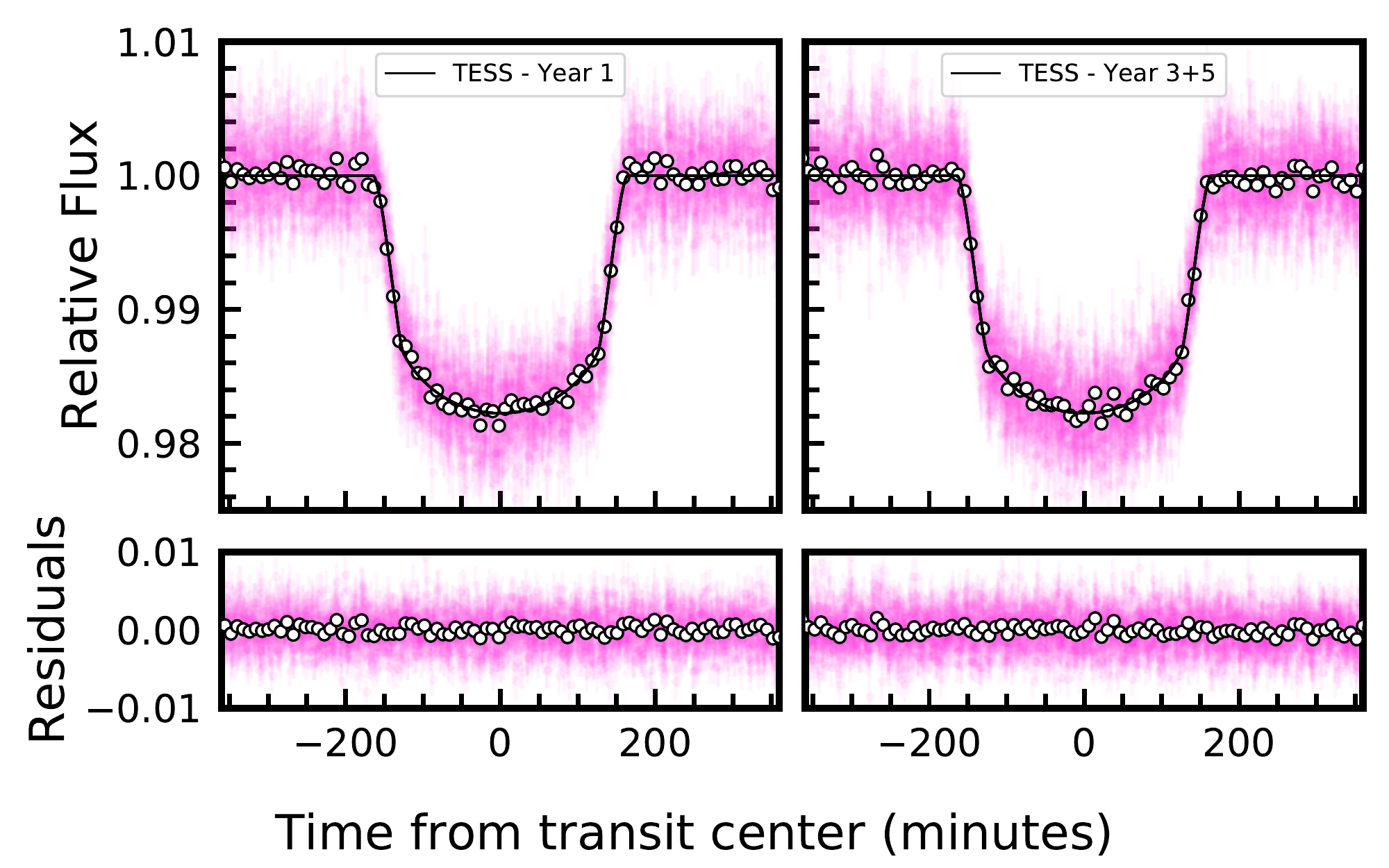}
\caption{Light curves of TOI-216\,c divided between year 1 (left) and year 3 onward (right)} of \textit{TESS} observations, phase aligned to the center of the transit. The best fitting model is shown in black for each individual transit. Binned data for each time interval is shown as white points.  
\label{fig:c_light_curves}
\end{figure}

The fitted light curves are displayed in Figures \ref{fig:b_light_curves} and \ref{fig:c_light_curves}. It is particularly clear for TOI-216\,b that the transits are deeper and wider in year 3 of the data. This is due to the impact parameter decreasing so that the planet is closer to fully transiting the star, crossing a longer chord and blocking more light. In the two transits from year 5, TOI-216\,b is fully transiting. The stellar and planetary parameters derived from the light curve fit are displayed in Table \ref{tab:light_curve_fitting}. As we are investigating changes to the impact parameter of the planets, we chose to only use \textit{TESS} data in order to provide a consistent dataset with respect to photometric noise properties and instrumental bandpass.

\begin{deluxetable}{lcc}[th]
\tablecaption{Planetary Parameters Derived from Light Curve Fit\label{tab:light_curve_fitting}}
\tablewidth{0pt}
\tablehead{
\colhead{Parameter} & \colhead{Value} & \colhead{68\% Confidence Interval}
}
\startdata
$M_\star~(M_\sun)$ & $0.763$ & $\pm0.021$\\
$R_\star~(R_\sun)$ & $0.757$ & $\pm0.007$\\
$\rho_\star~(\rho_\sun)$ & $1.76$ & $\pm0.06$\\
$u_1$ & $0.48$ & $\pm0.06$\\
$u_2$ & $0.04$ & $\pm0.13$\\
\\
$R_b/R_\star$ & $0.0949$ & $^{+0.0021}_{-0.0019}$\\
$R_c/R_\star$ & $0.1222$ & $\pm0.0008$\\
$R_b~(R_{\earth})$ & $7.84$ & $^{+0.21}_{-0.19}$\\
$R_c~(R_{\earth})$ & $10.09$ & $^{+0.15}_{-0.13}$\\
\\
$\overline{b}_{b}$ & $0.946$ & $^{+0.011}_{-0.010}$\\
$\overline{b}_{c}$ & $0.19$ & $^{+0.06}_{-0.09}$\\
$\gamma_b$ & $0.024$ & $^{+0.006}_{-0.005}$\\
$\gamma_c~(\text{95\% upper limit})$ & $<0.0006$\\
\enddata

\end{deluxetable}

\section{Analysis} \label{sec:analysis}

We jointly fit the transit times and impact parameters for each planet using the python \texttt{TTVFast} package \citep{Deck_2014}. For each planet, we assign a mass $M$, period $P$, eccentricity $e$, inclination $i$, longitude of the ascending node $\Omega$, argument of periapsis $\omega$, and mean longitude $\lambda$. We fix the longitude of the ascending node for the inner planet, TOI-216\,b,  as $\Omega_b = 0$, as only the difference in angle between the nodes of the two planets can be inferred. To account for the possibility that the uncertainties in the times of transit have been underestimated, we fit a  ``jitter'' parameter that is added in quadrature to the uncertainties in transit times themselves. We also allow the stellar mass and radius to vary following Gaussian priors derived from the analysis described in Section \ref{sec:data}. For all other parameters we apply a uniform prior. We set the time of the initial epoch to 1325.31 days, commensurate to the analysis in \citet{Dawson_2021}. \texttt{TTVFast} returns calculated times of transits, in addition to the on-sky separation between the planet and the center of the star, allowing the impact parameter to be determined. We sample with \texttt{emcee} \citep{Foreman_Mackey_2013}, an implementation of the affine invariant MCMC sampler of \citet{MCMC}, with the maximum-likelihood sample used as starting conditions for an N-body integration. The sampling used 500 walkers and 15,000 steps, with the first 10,000 discarded as burn-in.

The planetary parameters and orbital elements derived from the fitting are presented in Table \ref{tab:planets}. We calculate the mutual inclination $i_{\text{mut}}$ following
\begin{equation}
\cos{i_{\text{mut}}} = \cos{i_b}\cos{i_c} + \sin{i_b}\sin{i_c}\cos{(\Omega_b-\Omega_c)},
\label{eq:mutual_inclination}
\end{equation}
 and determine the density of the planet $\rho$ is from the mass and radius as inferred by the \texttt{TTVFast} fitting and light curve fitting respectively. The ``jitter'' term added an extra 1.7 minutes of uncertainty in quadrature to the transit times.

\begin{deluxetable}{lcc}
\tablecaption{Planetary Parameters at 1325.31 days for TOI-216\,b and TOI-216\,c\label{tab:planets}}
\tablewidth{0pt}
\tablehead{
\colhead{Parameter} & \colhead{Value} & \colhead{68\% Confidence Interval}
}
\startdata
$M_\star~(M_\sun)$ & $0.763$ & $\pm0.021$\\
$R_\star~(R_\sun)$ & $0.757$ & $\pm0.007$\\
\\
$M_b~(M_{\text{Jup}})$ & $0.0554$ & $\pm0.0020$\\
$P_b$ (days) & $17.0988$ & $^{+0.0007}_{-0.0006}$\\
$e_b$ & $0.1593$ & $^{+0.0021}_{-0.0019}$\\
$i_b~(\degr)$ & $88.554$ & $\pm0.020$\\
$\Omega_b~(\degr)$ & $0$\\
$\omega_b~(\degr)$ & $292.0$ & $\pm0.7$\\
$\lambda_b~(\degr)$ & $82.18$ & $^{+0.17}_{-0.16}$\\
\\
$M_c~(M_{\text{Jup}})$ & $0.525$ & $\pm0.019$\\
$P_c$ (days) & $34.5508$ & $\pm0.0003$\\
$e_c$ & $0.009$ & $^{+0.004}_{-0.003}$\\
$i_c~(\degr)$ & $89.801$ & $\pm0.019$\\
$\Omega_c~(\degr)$ & $-0.80$ & $\pm0.12$\\
$\omega_c~(\degr)$ & $236$ & $^{+7}_{-11}$\\
$\lambda_c~(\degr)$ & $27.50$ & $\pm0.18$\\
\\
$i_{\text{mut}}~(\degr)$ & $1.49$ & $\pm0.07$\\
$\rho_b~(\text{g cm}^{-3})$ & $0.201$ & $\pm0.017$\\
$\rho_c~(\text{g cm}^{-3})$ & $0.89$ & $\pm0.05$\\
\enddata
\end{deluxetable}

\begin{figure}[tb!]
\includegraphics[width=\columnwidth]{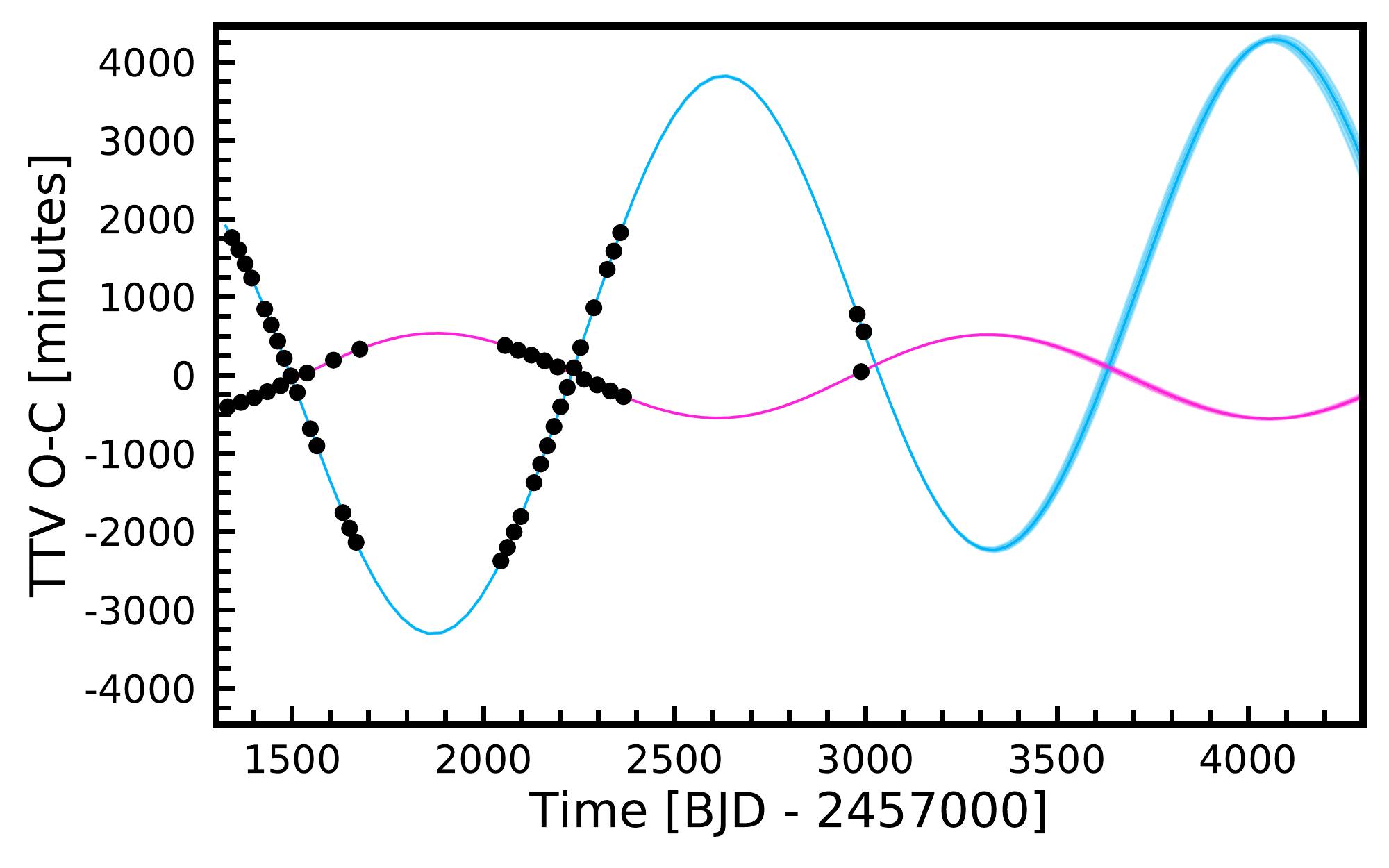}
\caption{Observed TTVs for TOI-216\,b (blue) and c (pink) shown as points. Error bars for b ($\sim$ 3 minutes) and c ($\sim$ 1 minute) are too small to be seen at this scale. Narrow shaded regions indicate 68\% and 95\% credible intervals of fitted models . The TTVs are anticorrelated with TOI-216\,b having a larger amplitude due to its lower mass and higher eccentricity. The TTVs change quasi-sinusoidally over a period of $\sim$1,500 days.}
\label{fig:both_ttvs}
\end{figure}

\section{Results and Discussion} \label{sec:results}

The two planets have very large TTV amplitudes: the TTV amplitude of TOI-216\,b is approximately 3,200 minutes, and the amplitude for TOI-216\,c is approximately 540 minutes. We show the measured TTVs for the two planets in Figure \ref{fig:both_ttvs}. The TTVs are anticorrelated and vary with a period of 1,500 days due to resonant libration. There is an additional component over 8,900 days due to apsidal precession of 1,100 minutes for TOI-216\,b and 10 minutes for TOI-216\,c.

\subsection{Change in Impact Parameter of Planet b}

\begin{figure}[tb!]
\includegraphics[width=\columnwidth]{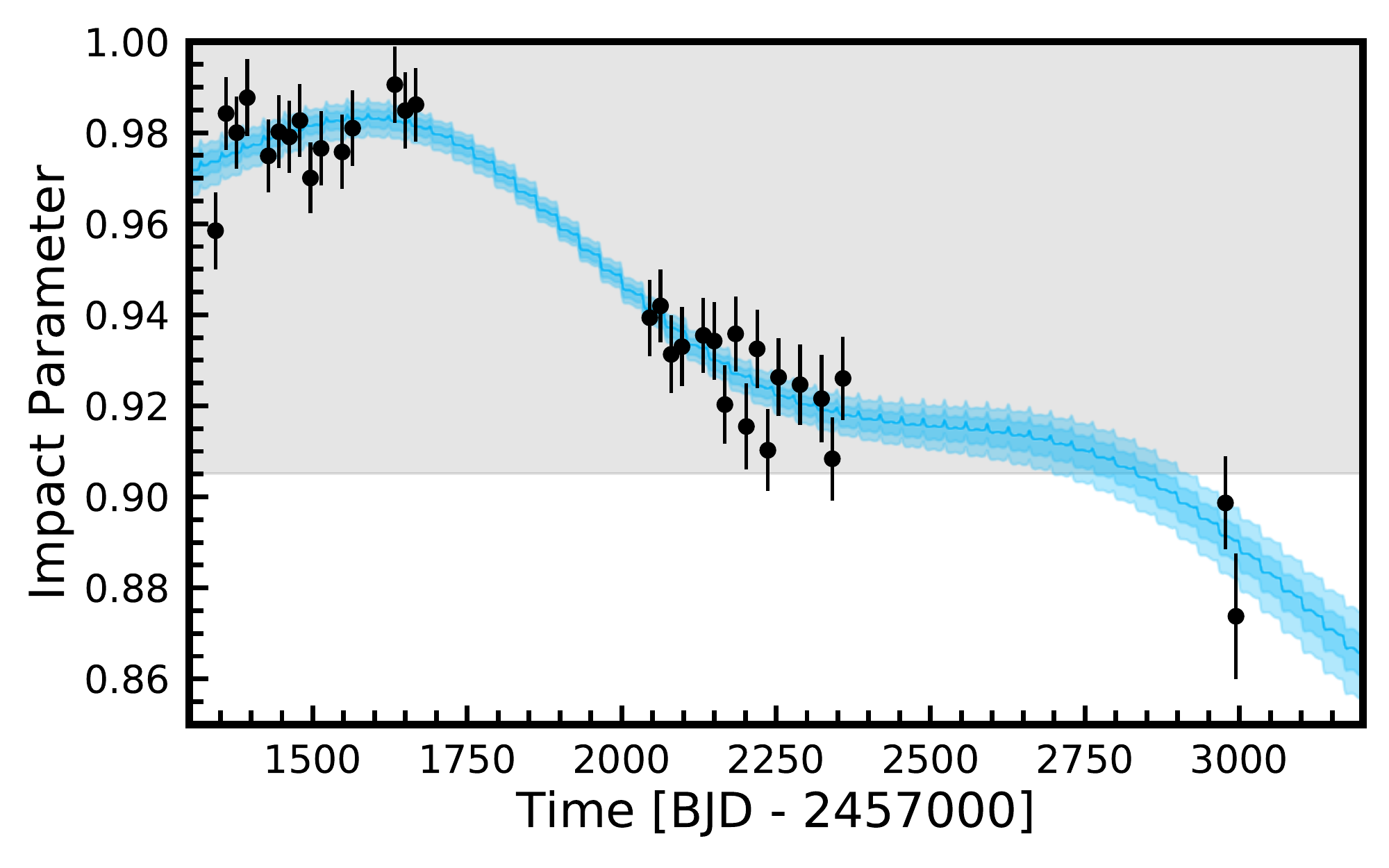}
\caption{Measured impact parameters from the light curve fitting alone for each transit of TOI-216\,b are shown as black points. The spread of dynamical models fit to the data is shown in blue, while the shaded regions mark 68\% and 95\% of n-body simulations. The light gray region indicates a grazing transit. The impact parameter for TOI-216\,b has changed from grazing to fully transiting over the \textit{TESS} observing baseline.}  
\label{fig:b_impact_parameter}
\end{figure}

\begin{figure}[tb!]
\includegraphics[width=\columnwidth]{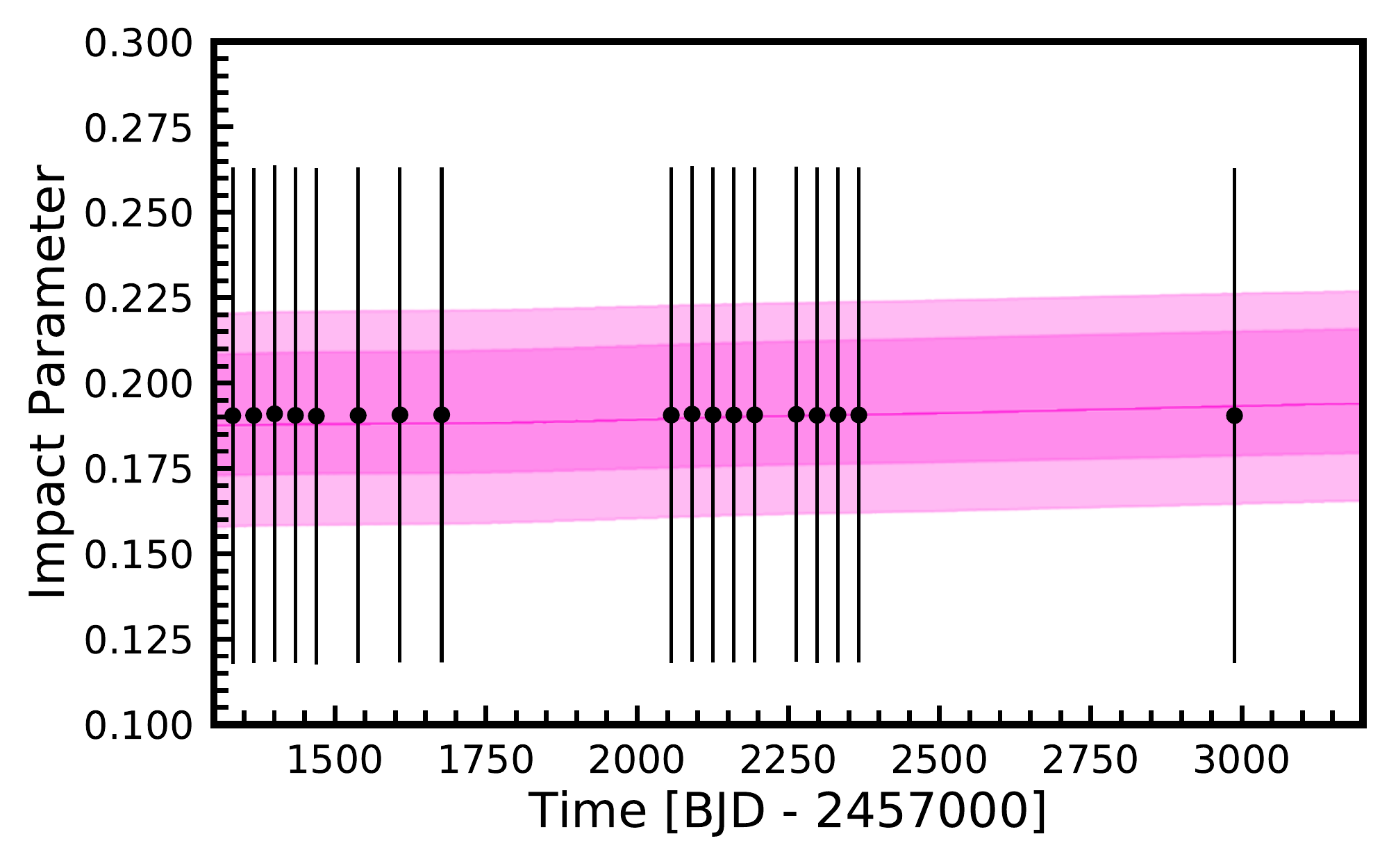}
\caption{Measured impact parameters from the light curve fitting alone for each transit of TOI-216\,c are shown as black points, following the approach of \citet{Dawson_2020}. The spread of dynamical models fit to the data is shown in pink, while the shaded regions mark 68\% and 95\% of n-body simulations. The individual transit impact parameters have highly correlated measurement uncertainties. The impact parameter for TOI-216\,c has no significant change over the \textit{TESS} observation baseline.}  
\label{fig:c_impact_parameter}
\end{figure}

We show a comparison between the measured impact parameters for both planets and model predictions for in Figures \ref{fig:b_impact_parameter} and \ref{fig:c_impact_parameter}. TOI-216\,b shows a large change in impact parameter between year 1 and year 3, changing from around 0.98 to 0.92 and continuing to decrease to 0.88 in year 5. As the radius ratio of planet b to the star is $0.0949^{+0.0021}_{-0.0019}$, transits with impact parameter larger than 0.91 are grazing, while impact parameters below that level suggest fully transiting events.In year 1 and 3  the transits are grazing, and the impact parameter is relatively constant throughout year 1, which is why no change was detected by \citet{Dawson_2020}. At the start of year 5 we infer the outer planet has a 95\% probability of being fully transiting, which increases throughout the year. We conclude that in these data the impact parameter has decreased to a point where TOI-216\,b is now fully transiting.

TOI-216\,c exhibits little change in impact parameter over the observed period, with an upper limit on the scale of the change calculated from the Cauchy distribution placed at 0.0006. We find from the light curve fit there is no detectable change across the observation period, although there is a large uncertainty in the mean impact parameter. There is greater uncertainty in measuring the transit impact parameter when it is lower \citep{Pal_2008}. From fitting the model to transit times and impact parameters we find that the spread of possible impact parameters is narrower than from the light curve alone. The model suggests that there should be an increase in impact parameter of $\sim 0.01$ $R_\star$ across the 4 year observing baseline. The changes in impact parameter are due to interactions between the two planets, so the more massive planet c is expected to be perturbed less than the lighter planet b.

\begin{figure}[tb!]
\includegraphics[width=\columnwidth]{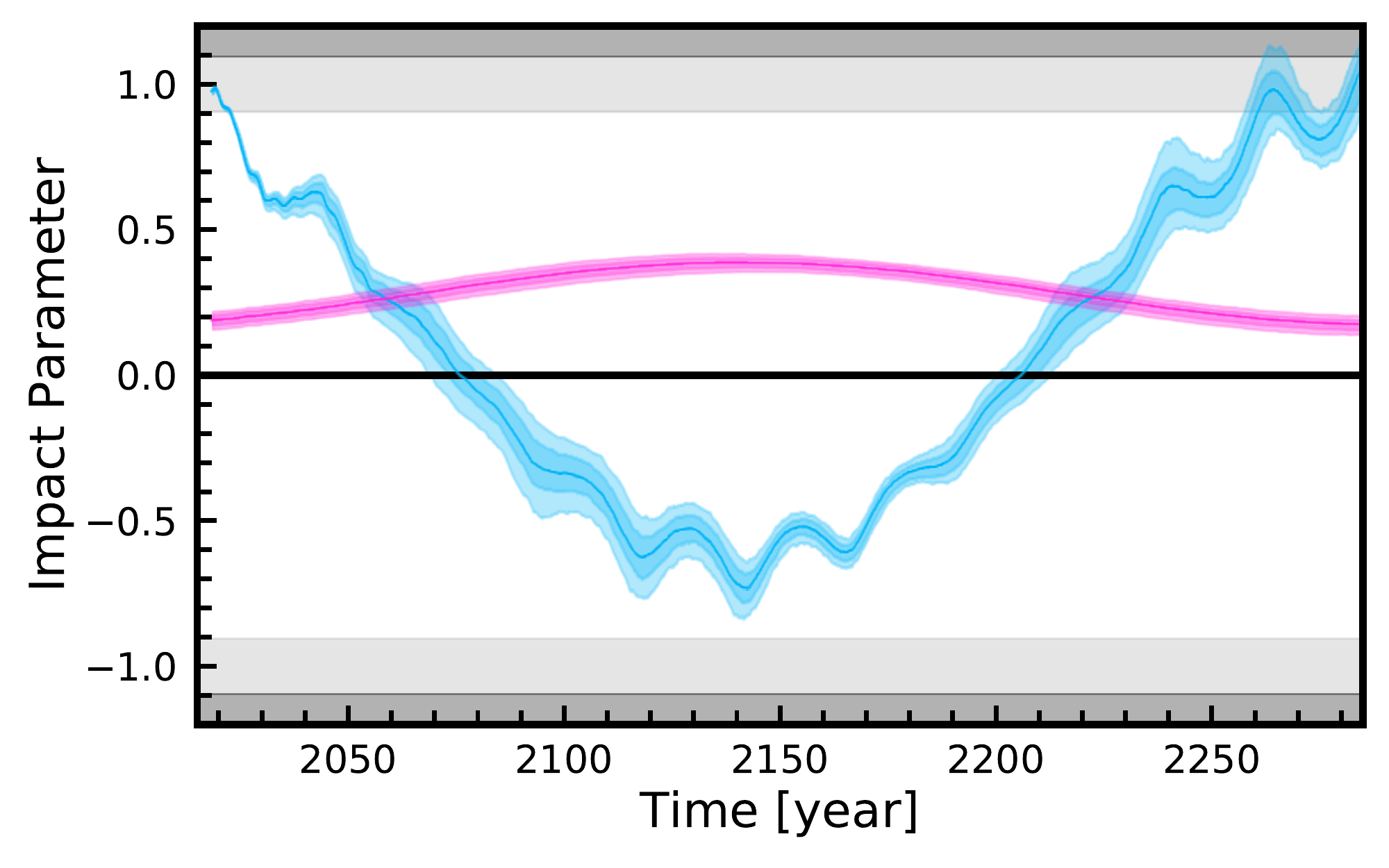}
\caption{Impact parameters for TOI-216\,b in blue and TOI-216\,c in pink, with shaded regions marking 68\% and 95\% of fitted models. A negative impact parameter is used as TOI-216\,b crosses over the face of the star to the opposite side relative to TOI-216\,c. The light gray shaded region marks a grazing transit for TOI-216\,b, the dark gray region indicates the planet has stopped transiting entirely. TOI-216\,b will fully transit for over 200 years, TOI-216\,c is always fully transiting.}  
\label{fig:both_impact_parameter}
\end{figure}

We project the future impact parameter of both planets in Figure \ref{fig:both_impact_parameter}. It can be seen that the period over which \textit{TESS} observed the system covers a fortuitous time, as it captures TOI-216\,b progressing from partially to fully transiting. This planet will continue to fully transit the star for the next 100 years. In doing so it will cross over to the opposite side of the star than TOI-216\,c. The outer planet's impact parameter does not have as drastic a change, always fully transiting and staying on one side of the center of the star. The timescale over which the impact parameter changes appears to be approximately 300 years, driven by a slow anticorrelated variation in the inclination of the planets. Shorter timescale variations are caused by apsidal precession, cyclic changes to the argument of periapsis of the inner planet over an 8,900 day period. As the orbit of the outer planet is nearly circular ($e = 0.009^{+0.004}_{-0.003}$), this effect is much smaller, so no obvious short term variation is seen in its impact parameter.

TOI-216 joins a small number of systems with observed transit duration variations, including Kepler-46 \citep{Dawson_2020}, Kepler-108 \citep{Mills_Fabrycky_2017}, Kepler-448 and Kepler-693 \citep{Masuda_2017}, K2-146 \citep{Hamann_2019}, and 15 KOIs identified by \citet{Shahaf_2021}. Future observations of known exoplanet systems will increase this population, with longer-term dynamics and the presence of non-transiting planets able to be explored in other Kepler systems \citep{Christ_2019, Goldberg_2019}.

\subsection{Synodic Chopping}

The n-body integrations of the TOI-216 system reveal synodic chopping in the transit times of planet b. As described by \citet{Deck_2014}, chopping occurs where the size of the TTVs jumps between successive transits. In the case of this system, the chopping signal is much smaller than the TTVs: it is on the order of a few minutes at its largest point during year 3 and around 1 minute in year 1. Planet c exhibits no significant chopping in its transit times due to the 2:1 orbital resonance resulting in planet b being in close to the same relative location each orbit.

The close encounters of the planets as they pass each other also affect other parameters of the orbits, although only impact parameter can be easily measured. Planet b has minor chopping-like changes to its impact parameter, but the scale is significantly smaller than the uncertainties in the impact parameter fit. A step-like signal can be seen in the modeled solutions for the impact parameter in Figure \ref{fig:b_impact_parameter}, particularly when the impact parameter is varying rapidly. The size of these steps is smaller than the uncertainty in the impact parameter, so it is unlikely that these could be directly measured with data from \textit{TESS}.

\subsection{Limits on a Third Planet}

The existence of a third planet had previously been invoked as one possible method of exciting the eccentricity of TOI-216\,b \citep{Dawson_2021}. Although subsequent modeling of planet migration found that the eccentricity of b and libration amplitude can be explained without the need for a third planet \citep{Nesvorny_2022}, here we consider the possibility of a third planet in the system, in either a 2:1 resonance inside planet b’s orbit ($P \approx 8.5$ days), or a 1:2 resonance outside planet c’s orbit ($P \approx 69$ days).

In both cases sampling favored planets with masses smaller than those of either known planet; more massive planets affect the structure of the TTV signal compared to the observed timing variations. We find a 95\% upper mass limit on an interior planet at half the period of planet b of 1.04 $M_{\earth}$; for an outer planet at twice the period of planet c, the limit is 11.8 $M_{\earth}$. In comparing the model to the data, we find no justification to include a third planet given the current data. The presence of a third planet at either location does not change the inferred masses of TOI-216\,b and c by a significant amount.

\subsection{Density of Planet b}

The radius of TOI-216\,b is now well characterized as it is fully transiting, with a radius of $7.84^{+0.21}_{-0.19}$ $R_{\earth}$, improving on the precision of previous estimates by a factor of 10. The radius of planet c is determined to be $10.09^{+0.15}_{-0.13}$ $R_{\earth}$, which is consistent with previous work. Planet b has a mass of $17.6\pm0.6$ $M_{\earth}$ and planet c has a mass of $167\pm6$ $M_{\earth}$. From this we calculate bulk densities of $0.201\pm0.017$ g cm$^{-3}$ and $0.89\pm0.05$ g cm$^{-3}$ for planets b and c respectively. The masses of the two planets are highly correlated, with a mass ratio between the planets $M_c/M_b$ of $9.48\pm0.02$. The radii are also somewhat correlated, the radius ratio $R_c/R_b$ is $1.29\pm0.03$, resulting in a density ratio of $\rho_c/\rho_b$ of $4.4\pm0.3$.

TOI-216\,b has a mass similar to that of Neptune, but a radius nearly twice as large, giving it a much lower density. Planet b's density is low for its mass but not unusual, as planets of similar mass and radius exist, such as Kepler-18\,d \citep{Cochran_2011}, which has a very similar mass, radius, density and orbital period. Kepler-18\,d has a mass of 16.4 $M_\earth$ and a radius of 6.98 $R_\earth$, so a density of 0.27 g cm$^{-3}$ and a 14.86 day orbit. This planet is the outer planet in a 2:1 resonance with Kepler-18\,c, a slightly smaller puffy Neptune. The radius, mass, and density of TOI-216\,c are between that of Saturn and Jupiter. Planet c's mass and radius are typical for a planet of its size. 

Kepler-117 is a system with a similar arrangement of planets. Kepler-117\,b has a mass of 29.9 $M_\earth$ and a radius of 8.06 $R_\earth$ giving it a low density of 0.30 g cm$^{-3}$, while Kepler-117\,c has a mass of 585 $M_\earth$ and a radius of 12.3 $R_\earth$, so a density of 1.74 g cm$^{-3}$ \citep{Bruno_2015}. The presence of a low density super-puff interior to a warm Jupiter resembles the TOI-216 system, a key difference is that with orbital periods of 18.8 and 50.8 days, Kepler-117 is not in a resonance and exhibits only small TTVs. Another similar system is Kepler-9, which has a pair of warm, puffy outer planets b and c with radii of 7.91 and 7.76 $R_\earth$ and densities of 0.495 and 0.362 g cm$^{-3}$ respectively \citep{Freudenthal_2018}. Kepler-9\,b and c are close to a 2:1 resonance with periods of 19.247 and 38.944 days, and exhibit large TTVs. This system is also predicted to show a change in the inclination and impact parameter of its planets over time, which has been detected in transit duration variations \citep{Shahaf_2021}.

We find that the uncertainty in the mass and radius measurements of the planets is primarily due to uncertainty in the mass and radius of the star. The planet-to-star ratios for mass have $1\sigma$ uncertainties of $2.4\%$, while the uncertainty in the stellar mass reported by \citet{Dawson_2021} is $2.8\%$. We find that the radius ratios for the planets have uncertainties of $2.1\%$ and $0.65\%$ for b and c respectively, while the stellar radius has an uncertainty of $0.92\%$. The dynamics of this system are such that these parameters for the planets are very well characterized. Therefore the benefits of including additional data such as ground-based photometry, or radial velocity observations are limited. The most significant parameters are already well known: characterization of TOI-216 itself is the limiting factor in refining the size of the planet.

Super-puffs are planets with uncommonly low densities, typically below 0.3 g cm$^{-3}$ \citep{Liang_2021}, like TOI-216\,b. Many super-puffs have masses comparable to super-Earths \citep{Lee_2016}. However, the larger 12.0 $M_{\earth}$ HIP 41738\,f \citep{Belkovski_2022}, 15.0 $M_{\earth}$ Kepler-90\,g \citep{Liang_2021}, 16.4 $M_{\earth}$ Kepler-18\,d \citep{Cochran_2011} and 30.5 $M_{\earth}$ WASP-107\,b \citep{Piaulet_2021} have similar masses and comparable low densities to TOI-216\,b.

One explanation for the inflated radii of super-puffs is the presence of high-altitude hazes due to the outflow of the atmosphere \citep{Gao_2020}. A similar explanation utilizes the formation of dust grains in the outflow to  contribute to increasing the apparent radius of the planet \citep{Ohno_2021}. Both of these explanations predict a featureless transmission spectrum, which as been observed in Kepler-79\,d \citep{Chachan_2020}. These explanations are most effective at explaining super-puffs with a mass $<5$ $M_{\earth}$, so they may not explain the low density of the higher mass TOI-216\,b. The very low density and low latitude, placing it in the \textit{TESS} southern continuous viewing zone, may make this target an attractive follow-up target with JWST to probe super-puff atmospheres.

These low density planets are often observed in systems exhibiting TTVs. The masses and densities of planets found in TTV systems are systematically lower than those calculated using RV surveys \citep{Weiss_2014}. It is possible that this difference is due to the presence of super-puff planets in or near a mean motion resonance that migrated inwards to their current locations after forming further out in the disk \citep{Lee_2016}. Then low density planets are more likely to be found near a resonance and therefore will exhibit a larger TTV amplitude. The density of TOI-216\,b and the circular orbit of c are both suggestive of inwards migration in this way. 

Another explanation for the difference in trends between masses derived through TTVs and RVs is that for a given radius of planet RVs tend to detect higher mass planets, while TTVs have a more uniform sensitivity \citep{Steffen_2016}, which would amplify any potential physical differences between the samples. An analysis of the mass to radius ratios of planets found that the difference in masses can be attributed to differing sensitivities based on period \citep{Mills_Mazeh_2017}. For short period orbits TTVs and RVs produce similar results, while at longer periods RVs have a detection bias towards higher mass planets. For planets that have mass determinations from both TTVs and RVs, results are generally consistent with no clear bias of masses in one direction. Using TTVs, we find masses of TOI-216\,b and c that are consistent with those found by \citet{Dawson_2021}, where a joint transit and radial velocity fit was performed.

The tidal migration of the planets around TOI-216 has been explored by \citet{Nesvorny_2022}. They find that the resonant architecture of the system can be explained by TOI-216\,b having its migration stall at the inner edge of the protoplanetary disk, where the outer planet can then migrate inwards to capture the system into a 2:1 resonance. The process of Neptunes and sub-Neptunes becoming trapped at the inner edge of the disk was explored by \citet{Chrenko_2022}, using TOI-216\,b as a test case.

\subsection{Future Transits}

\begin{figure}[tb!]
\includegraphics[width=\columnwidth]{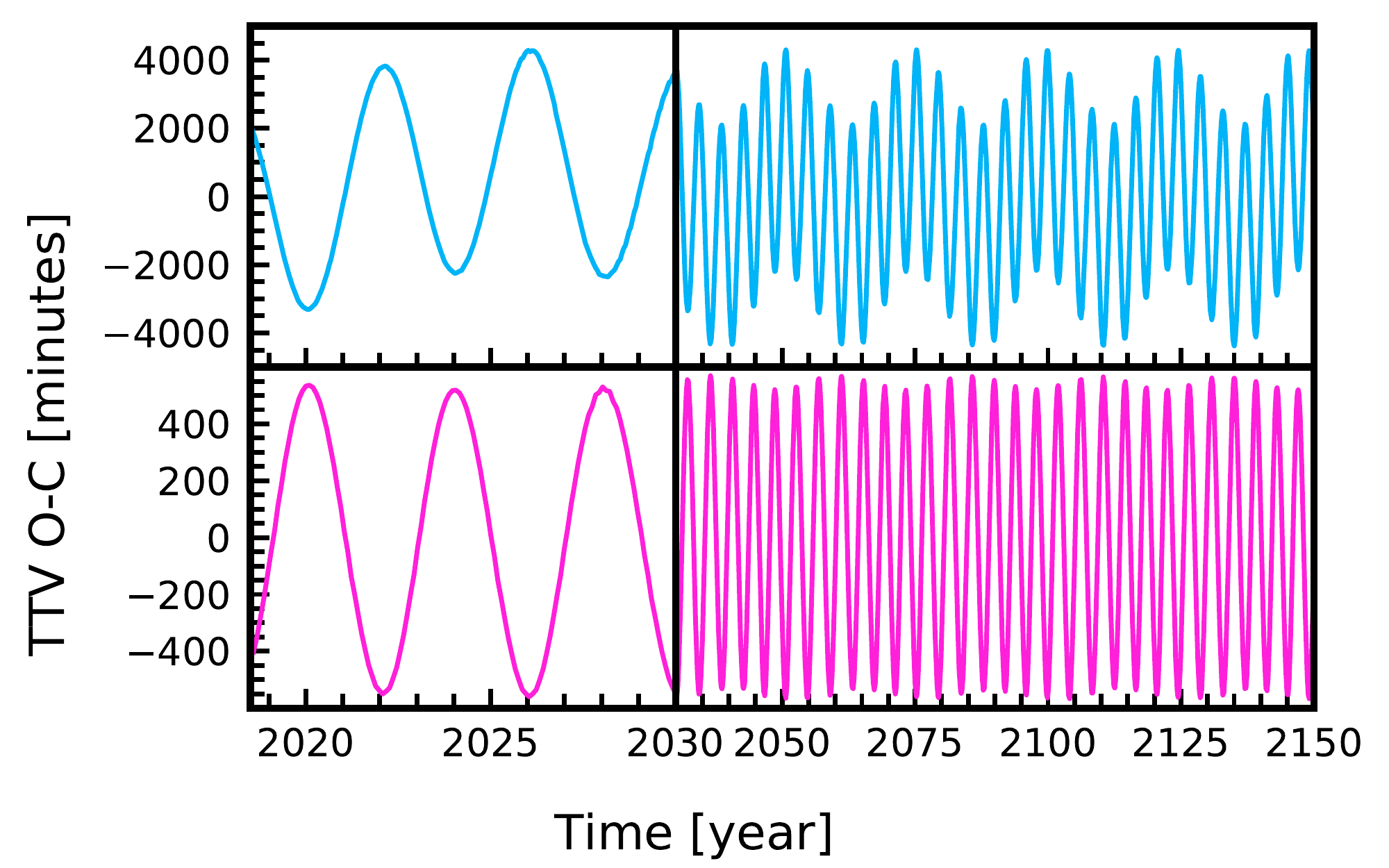}
\caption{Transit timing variations for TOI-216\,b (blue) and c (pink) over a 150 year period based on the maximum likelihood dynamical model of the system. The left panels highlight a shorter period of time to show a 1,500 day long periodicity in TTVs. The right panel shows a more slowly varying 8,900 day period in transit times.}  
\label{fig:both_ttvs_long}
\end{figure}

\begin{figure}[tb!]
\includegraphics[width=\columnwidth]{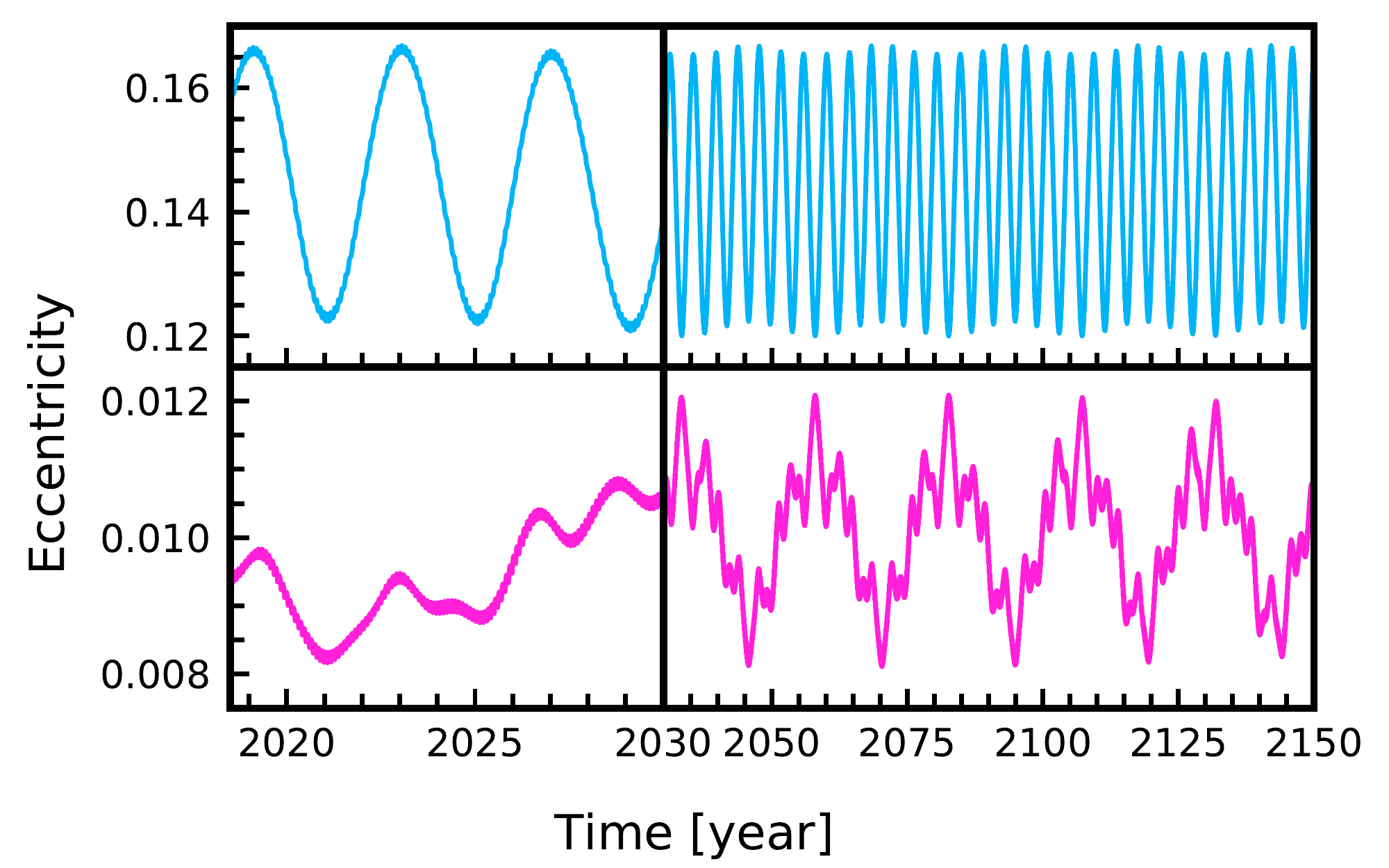}
\caption{Eccentricity variation for TOI-216\,b (blue) and c (pink) over a 150 year period based on the maximum likelihood dynamical model of the system. The left panel highlights a shorter period of time to show a 1,500 day long periodicity in the eccentricity of b. The right panel shows a more slowly varying 8,900 day period in eccentricity.}  
\label{fig:both_eccentricities}
\end{figure}

The projected transit timing variations for TOI-216\,b and c are shown in Figure \ref{fig:both_ttvs_long}, using the maximum likelihood fitted model. The eccentricity of the orbit of each planet is shown in Figure \ref{fig:both_eccentricities}.

Both the TTVs and eccentricity variations of each planet show a short term period of approximately 1,500 days, and a longer period of approximately 8,900 days. These TTVs are calculated using a time-averaged period of 17.21720 days for the inner planet, and 34.50113 days for the outer planet, giving a period ratio of 2.00388. Using Equation \ref{eq:super_period} this period ratio gives a super-period of 8,902 days, explaining the 8,900 day longer term period seen. The larger amplitude 1,500 day variation is due to the libration of the system around mean motion resonance. The osculating periods of the two planets vary between 17.05 to 17.40 days for b, and 34.44 to 34.56 days for c, with the osculating period ratio changing between 1.98 to 2.03. The long term TTV amplitude is more significant for the inner planet due to its lower mass.

\begin{figure}[tb!]
\includegraphics[width=\columnwidth]{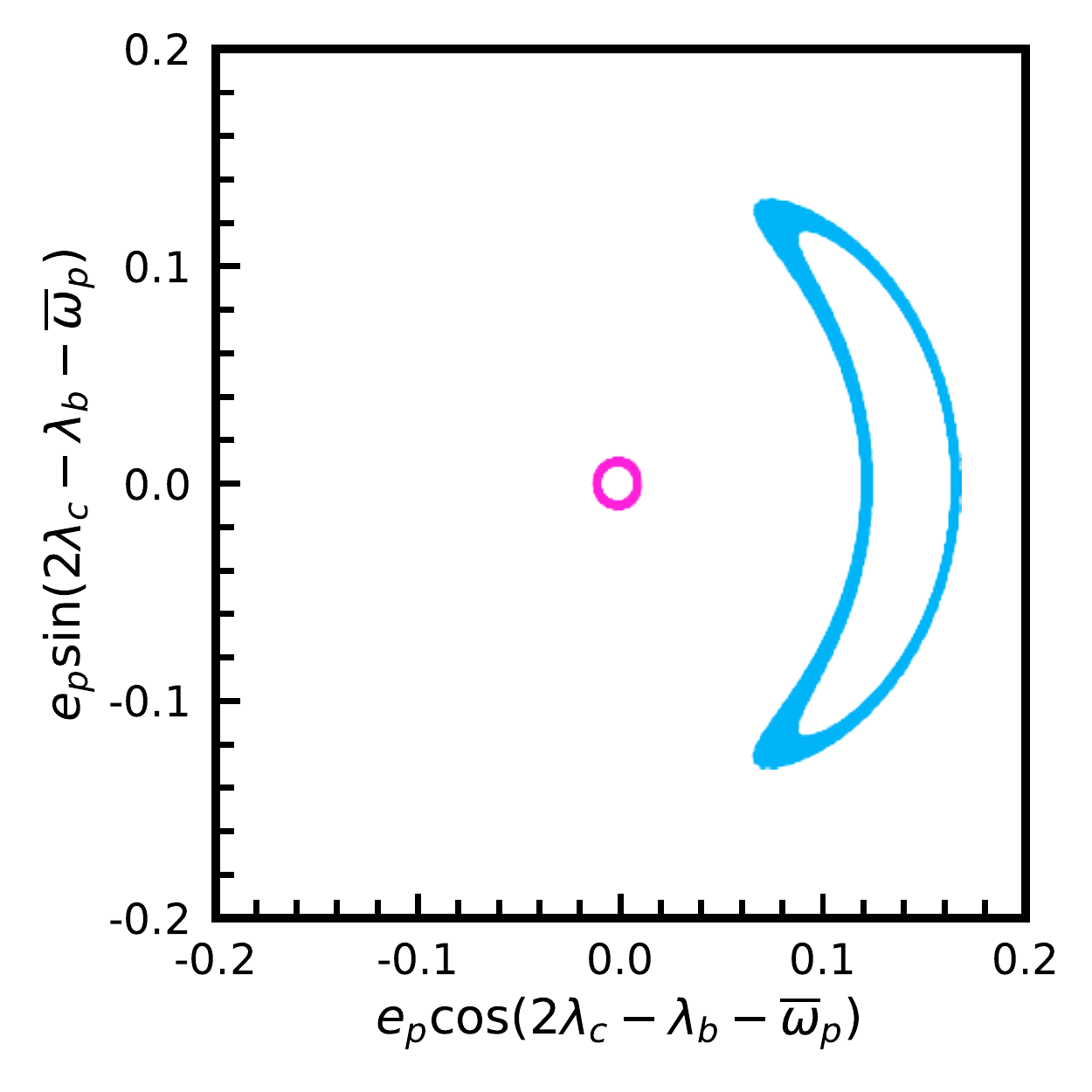}
\caption{The trajectory of the maximum likelihood dynamical model showing the libration of angles around a fixed point for TOI-216\,b (blue), and the circulating nature of TOI-216\,c (pink).}  
\label{fig:b_resonance}
\end{figure}

The 2:1 resonant angle $2\lambda_{c}-\lambda_{b}-\overline{\omega}_{b}$, where $\overline{\omega}$ is the longitude of periapsis, for TOI-216\,b librates around a fixed point. This is shown in Figure \ref{fig:b_resonance} where all sampled solutions librate in a 2:1 resonance. The inner planet librates over a timescale of 1,500 days. This resonant angle for TOI-216\,c circulates in all sampled solutions over 150 years.

\vspace{3em}

\subsection{Orbital Evolution}

The orbit of TOI-216\,b has a moderate eccentricity  of $0.1593^{+0.0021}_{-0.0019}$, while TOI-216\,c has a nearly circular orbit with an eccentricity of only $0.009^{+0.004}_{-0.003}$. The eccentricity of planet b varies between 0.120 and 0.167 over the 1,500 day libration timescale. Planet c's eccentricity varies more slowly, with the largest amplitude effect occurring over a 8,900 day period. The low eccentricity of TOI-216\,c is consistent with in situ formation or resonance capture migration, while high-eccentricity migration seems unlikely.

\begin{figure}[tb!]
\includegraphics[width=\columnwidth]{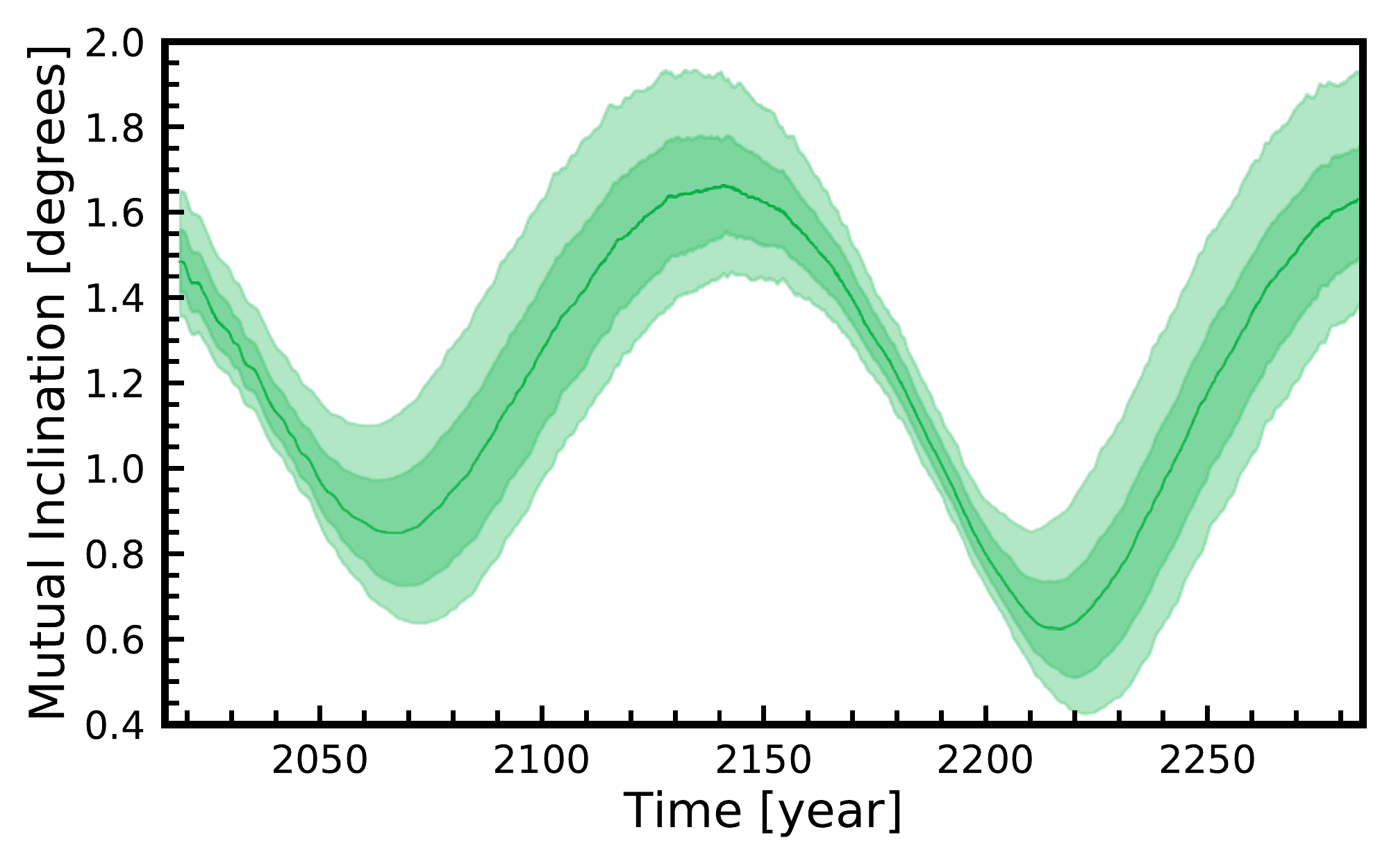}
\caption{The mutual angle between the orbital planes of TOI-216\,b and c with shaded regions marking 68\% and 95\% of fitted models. The changes in mutual inclination are primarily due to changes in the sky-plane inclination of the two planets, reaching minima where the impact parameters coincide.}  
\label{fig:mutual_inclination}
\end{figure}

The planets exhibit a mutual inclination of $1\fdg49\pm0.07$ which is observed in the difference in measured impact parameters. The torque exerted due to the mutual inclination causes the observed sky-plane inclinations of the two planets to change over time, driving the change in impact parameter seen in the transits of TOI-216\,b. The mutual inclination therefore changes over time, as seen in Figure \ref{fig:mutual_inclination}. The mutual inclination is smallest when the sky-plane inclinations, and so the impact parameters, of the two planets are the same. At these minima the mutual inclination is nearly equal to the difference in the longitudes of the ascending nodes of the two planets, $\Delta\Omega=\Omega_b-\Omega_c$. Within the next $\sim$300 years, the maximum mutual inclination between the orbital planes of the two planets is $1\fdg67\pm0.12$ and the minimum is $0\fdg62\pm0.11$, demonstrating a small misalignment between the planets.

Our model predicts that future transits will continue to show impact parameter changes. Transits of TOI-216\,b will show a dramatic decrease in impact parameter over the next decade, which will be observable as a lengthening of the transit duration. The impact parameter of TOI-216\,c slowly increases over the same timescale. The changes to the more massive planet c's impact parameter will be smaller, precluding the possibility of any syzygies in the system in the next few decades.

\subsection{Comparison with Previous Work}

We compare the values found in this paper to those most recently published for TOI-216 in \citet{Dawson_2021}. The stellar parameters used by \citet{Dawson_2021} were a mass of 0.77$\pm$0.03 $M_\sun$ and a radius of 0.748$\pm$0.015 $R_\sun$, where the uncertainties were not included in the dynamical fitting. These values were used as the prior for fitting the light curve, which found a best fit for a lower mass and larger radius, both have overlapping uncertainty. The radii and masses determined for the planets are consistent, with slight improvements in uncertainties. The exception is for the radius of TOI-216\,b, as we have greatly reduced the uncertainty in the measurement.

TOI-216 joins a collection of other systems with super-puff planets, such as Kepler-18 \citep{Cochran_2011}, Kepler-79 \citep{Jontof-Hutter_2014, Chachan_2020}, Kepler-90 \citep{Liang_2021} and Kepler-117 \citep{Bruno_2015}. The low density planets in these systems often have factors in common: they all have low orbital eccentricities, while most have periods that are close to an integer ratio with another planet in the system, leading to significant TTVs. Often, they are in or near resonance with a Jupiter analog. Kepler-117 does not show significant TTVs, but does feature an inner super-puff and a larger, Jupiter-mass outer planet, much like TOI-216. While TOI-216\,b has a moderately eccentric orbit, TOI-216\,c has an orbit that is very circular. The circular orbits of many of these planets and being in or close to a mean motion resonance suggests these super-puffs may have undergone disk migration to their current locations, as suggested by \citet{Lee_2016}. The commonalities of the orbits of these planets suggest they are the result of a shared formation mechanism, rather than a shorter-term observational effect such as the presence of rings, as explored by \citet{Piro_2020}.

\begin{figure}[htb!]
\includegraphics[width=\columnwidth]{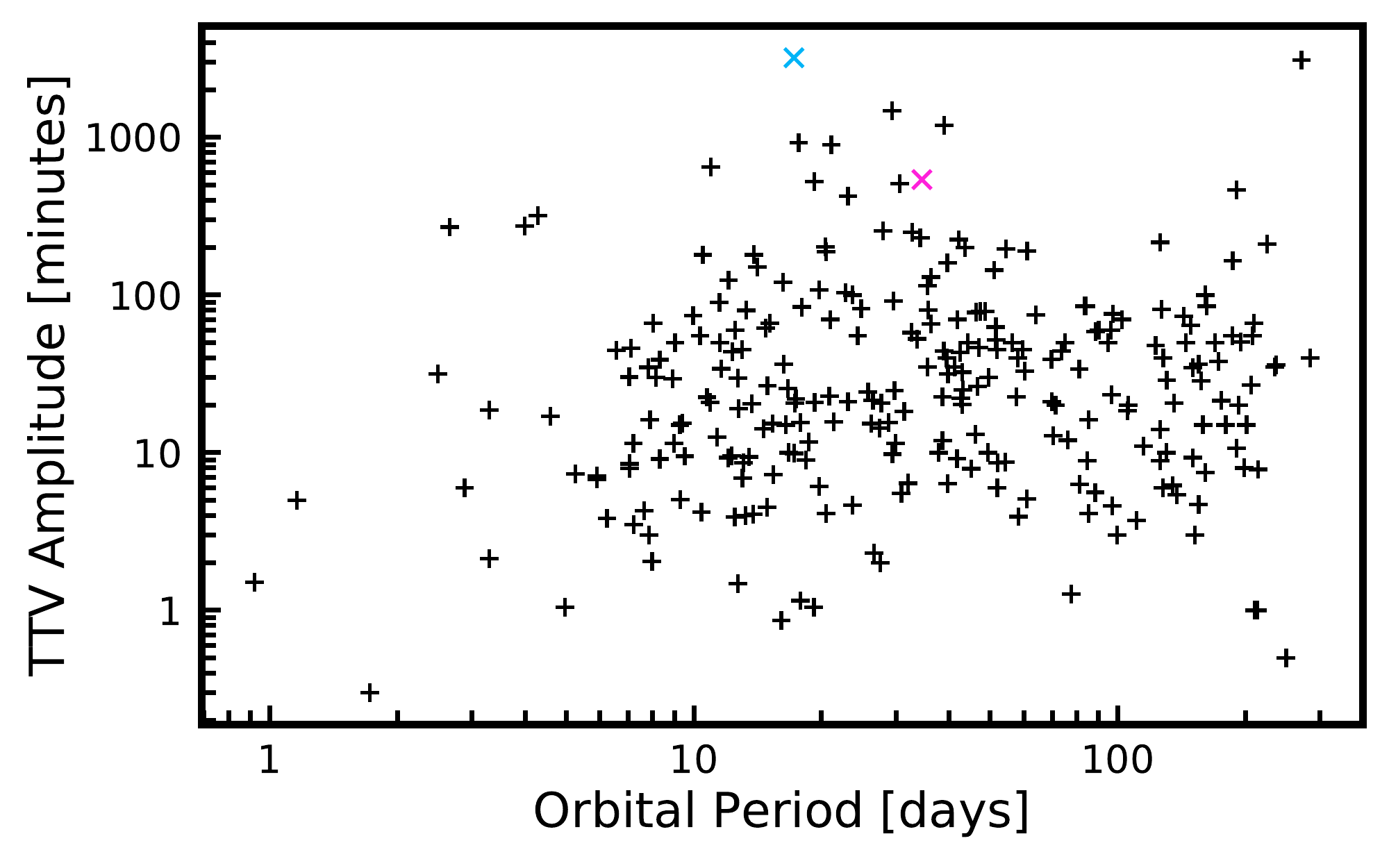}
\caption{The amplitude of TTV oscillations for planets in known TTV systems against the average orbital period of the planet. TOI-216\,b is marked in blue, TOI-216\,c is marked in pink}  
\label{fig:ttv_amplitudes}
\end{figure}

We compare the TTV amplitudes of TOI-216\,b and c to those of other known TTV systems in Figure \ref{fig:ttv_amplitudes}. In can be seen that TOI-216\,b and TOI-216\,c have some of the highest TTV amplitudes of all known planets. If the TTV amplitude is calculated as a fraction of the orbital period, TOI-216\,b is the highest.

\section{Conclusions} \label{sec:conclusions}

The extreme TTVs of the planets in the TOI-216 system are used to precisely characterize the orbits of these planets. TOI-216\,b has changed from a grazing transit to fully transiting, so we can now accurately measure its radius. TOI-216\,b is a puffy Neptune-mass planet ($17.6\pm0.6$ $M_{\earth}$), with a much larger radius that is now well constrained to $7.84^{+0.21}_{-0.19}$ $R_{\earth}$, TOI-216\,c is a warm Jupiter with a typical mass ($167\pm6$ $M_{\earth}$) and radius ($10.09^{+0.15}_{-0.13}$ $R_{\earth}$). We calculate bulk densities of $0.201\pm0.017$ g cm$^{-3}$ and $0.89\pm0.05$ g cm$^{-3}$ for planets b and c respectively. The mass and radius measurement precision is now limited by uncertainties in stellar parameters.

Future transits observed by \textit{TESS} will increase precision of orbital characteristics and radii measurements, but not by a large amount as the main uncertainty lies in the star itself. New transits will show changes in impact parameter that will confirm the mutual inclination between the planets, and the continued observation of TTVs will narrow uncertainties further.

TOI-216\,b is yet another super-puff found in a TTV system. Multiple low-density planets like these have been found on warm orbits that are close to or in mean motion resonances with other planets. They are often on low-eccentricity orbits, giving further weight to the ideas of \citet{Lee_2016} that super-puffs formed at larger distances from their host stars and migrated inwards, thus are captured into resonance, explaining their prominent TTVs. TOI-216 has a \textit{TESS} magnitude of 11.5 \citep{Dawson_2019}; its low density and bright host make it a suitable target for atmospheric characterization by JWST in order to better understand the composition of super-puffs.

\begin{acknowledgments}
We thank Rebekah Dawson (Penn State) for constructive discussions that improved the quality of this manuscript. We thank the anonymous referee for their prompt and helpful feedback.

This paper includes data collected by the \textit{TESS} mission. Funding for the \textit{TESS} mission is provided by the NASA's Science Mission Directorate. This research has made use of the Exoplanet Follow-up Observation Program website, which is operated by the California Institute of Technology, under contract with the National Aeronautics and Space Administration under the Exoplanet Exploration Program. This paper includes data collected by the \textit{TESS} mission, which are publicly available from the Mikulski Archive for Space Telescopes (MAST). The specific observations analyzed can be accessed via \dataset[https://doi.org/10.17909/9tff-qs49]{https://doi.org/10.17909/9tff-qs49}. STScI is operated by the Association of Universities for Research in Astronomy, Inc., under NASA contract NAS5–26555. Support to MAST for these data is provided by the NASA Office of Space Science via grant NAG5–7584 and by other grants and contracts.

This research made use of \texttt{exoplanet} \citep{exoplanet:joss,
exoplanet:zenodo} and its dependencies \citep{exoplanet:agol20,
exoplanet:arviz, exoplanet:astropy13, exoplanet:astropy18, exoplanet:luger18,
exoplanet:pymc3, exoplanet:theano}.

\end{acknowledgments}

\facilities{\textit{TESS}}

\software{
arviz \citep{exoplanet:arviz},
Astrocut \citep{astrocut},
Astropy \citep{exoplanet:astropy13, exoplanet:astropy18},
astroquery \citep{astroquery},
corner \citep{corner},
emcee \citep{Foreman_Mackey_2013},
exoplanet \citep{exoplanet:joss, exoplanet:zenodo},
Lightkurve \citep{lightkurve},
Matplotlib \citep{matplotlib},
NumPy \citep{numpy},
PyMC3 \citep{exoplanet:pymc3},
REBOUND \citep{rebound, reboundias15},
SciPy \citep{scipy},
starry \citep{exoplanet:luger18}
Theano \citep{exoplanet:theano},
TTVFast \citep{Deck_2014},
}

\bibliography{main}{}

\begin{thebibliography}{}
\expandafter\ifx\csname natexlab\endcsname\relax\def\natexlab#1{#1}\fi
\providecommand{\url}[1]{\href{#1}{#1}}
\providecommand{\dodoi}[1]{doi:~\href{http://doi.org/#1}{\nolinkurl{#1}}}
\providecommand{\doeprint}[1]{\href{http://ascl.net/#1}{\nolinkurl{http://ascl.net/#1}}}
\providecommand{\doarXiv}[1]{\href{https://arxiv.org/abs/#1}{\nolinkurl{https://arxiv.org/abs/#1}}}

\bibitem[{Agol \& Fabrycky(2018)}]{Agol_2018}
Agol, E., \& Fabrycky, D.~C. 2018, in Handbook of Exoplanets (Springer
  International Publishing), 797--816, \dodoi{10.1007/978-3-319-55333-7_7}

\bibitem[{{Agol} {et~al.}(2020){Agol}, {Luger}, \&
  {Foreman-Mackey}}]{exoplanet:agol20}
{Agol}, E., {Luger}, R., \& {Foreman-Mackey}, D. 2020, \aj, 159, 123,
  \dodoi{10.3847/1538-3881/ab4fee}

\bibitem[{{Agol} {et~al.}(2005){Agol}, {Steffen}, {Sari}, \&
  {Clarkson}}]{Agol_2005}
{Agol}, E., {Steffen}, J., {Sari}, R., \& {Clarkson}, W. 2005, \mnras, 359,
  567, \dodoi{10.1111/j.1365-2966.2005.08922.x}

\bibitem[{{Astropy Collaboration} {et~al.}(2013){Astropy Collaboration},
  {Robitaille}, {Tollerud}, {Greenfield}, {Droettboom}, {Bray}, {Aldcroft},
  {Davis}, {Ginsburg}, {Price-Whelan}, {Kerzendorf}, {Conley}, {Crighton},
  {Barbary}, {Muna}, {Ferguson}, {Grollier}, {Parikh}, {Nair}, {Unther},
  {Deil}, {Woillez}, {Conseil}, {Kramer}, {Turner}, {Singer}, {Fox}, {Weaver},
  {Zabalza}, {Edwards}, {Azalee Bostroem}, {Burke}, {Casey}, {Crawford},
  {Dencheva}, {Ely}, {Jenness}, {Labrie}, {Lim}, {Pierfederici}, {Pontzen},
  {Ptak}, {Refsdal}, {Servillat}, \& {Streicher}}]{exoplanet:astropy13}
{Astropy Collaboration}, {Robitaille}, T.~P., {Tollerud}, E.~J., {et~al.} 2013,
  \aap, 558, A33, \dodoi{10.1051/0004-6361/201322068}

\bibitem[{{Astropy Collaboration} {et~al.}(2018){Astropy Collaboration},
  {Price-Whelan}, {Sip{\H o}cz}, {G{\"u}nther}, {Lim}, {Crawford}, {Conseil},
  {Shupe}, {Craig}, {Dencheva}, {Ginsburg}, {VanderPlas}, {Bradley},
  {P{\'e}rez-Su{\'a}rez}, {de Val-Borro}, {Aldcroft}, {Cruz}, {Robitaille},
  {Tollerud}, {Ardelean}, {Babej}, {Bach}, {Bachetti}, {Bakanov}, {Bamford},
  {Barentsen}, {Barmby}, {Baumbach}, {Berry}, {Biscani}, {Boquien}, {Bostroem},
  {Bouma}, {Brammer}, {Bray}, {Breytenbach}, {Buddelmeijer}, {Burke},
  {Calderone}, {Cano Rodr{\'{\i}}guez}, {Cara}, {Cardoso}, {Cheedella},
  {Copin}, {Corrales}, {Crichton}, {D'Avella}, {Deil}, {Depagne}, {Dietrich},
  {Donath}, {Droettboom}, {Earl}, {Erben}, {Fabbro}, {Ferreira}, {Finethy},
  {Fox}, {Garrison}, {Gibbons}, {Goldstein}, {Gommers}, {Greco}, {Greenfield},
  {Groener}, {Grollier}, {Hagen}, {Hirst}, {Homeier}, {Horton}, {Hosseinzadeh},
  {Hu}, {Hunkeler}, {Ivezi{\'c}}, {Jain}, {Jenness}, {Kanarek}, {Kendrew},
  {Kern}, {Kerzendorf}, {Khvalko}, {King}, {Kirkby}, {Kulkarni}, {Kumar},
  {Lee}, {Lenz}, {Littlefair}, {Ma}, {Macleod}, {Mastropietro}, {McCully},
  {Montagnac}, {Morris}, {Mueller}, {Mumford}, {Muna}, {Murphy}, {Nelson},
  {Nguyen}, {Ninan}, {N{\"o}the}, {Ogaz}, {Oh}, {Parejko}, {Parley}, {Pascual},
  {Patil}, {Patil}, {Plunkett}, {Prochaska}, {Rastogi}, {Reddy Janga},
  {Sabater}, {Sakurikar}, {Seifert}, {Sherbert}, {Sherwood-Taylor}, {Shih},
  {Sick}, {Silbiger}, {Singanamalla}, {Singer}, {Sladen}, {Sooley},
  {Sornarajah}, {Streicher}, {Teuben}, {Thomas}, {Tremblay}, {Turner},
  {Terr{\'o}n}, {van Kerkwijk}, {de la Vega}, {Watkins}, {Weaver}, {Whitmore},
  {Woillez}, {Zabalza}, \& {Astropy Contributors}}]{exoplanet:astropy18}
{Astropy Collaboration}, {Price-Whelan}, A.~M., {Sip{\H o}cz}, B.~M., {et~al.}
  2018, \aj, 156, 123, \dodoi{10.3847/1538-3881/aabc4f}

\bibitem[{{Belkovski} {et~al.}(2022){Belkovski}, {Becker}, {Howe}, {Malsky}, \&
  {Batygin}}]{Belkovski_2022}
{Belkovski}, M., {Becker}, J., {Howe}, A., {Malsky}, I., \& {Batygin}, K. 2022,
  \aj, 163, 277, \dodoi{10.3847/1538-3881/ac6353}

\bibitem[{{Brasseur} {et~al.}(2019){Brasseur}, {Phillip}, {Fleming},
  {Mullally}, \& {White}}]{astrocut}
{Brasseur}, C.~E., {Phillip}, C., {Fleming}, S.~W., {Mullally}, S.~E., \&
  {White}, R.~L. 2019, {Astrocut: Tools for creating cutouts of TESS images},
  Astrophysics Source Code Library, record ascl:1905.007.
\newblock \doeprint{1905.007}

\bibitem[{{Bruno} {et~al.}(2015){Bruno}, {Almenara}, {Barros}, {Santerne},
  {Diaz}, {Deleuil}, {Damiani}, {Bonomo}, {Boisse}, {Bouchy}, {H{\'e}brard}, \&
  {Montagnier}}]{Bruno_2015}
{Bruno}, G., {Almenara}, J.~M., {Barros}, S.~C.~C., {et~al.} 2015, \aap, 573,
  A124, \dodoi{10.1051/0004-6361/201424591}

\bibitem[{{Chabrier} {et~al.}(2009){Chabrier}, {Baraffe}, {Leconte},
  {Gallardo}, \& {Barman}}]{Chabrier_2009}
{Chabrier}, G., {Baraffe}, I., {Leconte}, J., {Gallardo}, J., \& {Barman}, T.
  2009, in American Institute of Physics Conference Series, Vol. 1094, 15th
  Cambridge Workshop on Cool Stars, Stellar Systems, and the Sun, ed.
  E.~{Stempels}, 102--111, \dodoi{10.1063/1.3099078}

\bibitem[{{Chachan} {et~al.}(2020){Chachan}, {Jontof-Hutter}, {Knutson},
  {Adams}, {Gao}, {Benneke}, {Berta-Thompson}, {Dai}, {Deming}, {Ford}, {Lee},
  {Libby-Roberts}, {Madhusudhan}, {Wakeford}, \& {Wong}}]{Chachan_2020}
{Chachan}, Y., {Jontof-Hutter}, D., {Knutson}, H.~A., {et~al.} 2020, \aj, 160,
  201, \dodoi{10.3847/1538-3881/abb23a}

\bibitem[{{Chrenko} {et~al.}(2022){Chrenko}, {Chametla}, {Nesvorn{\'y}}, \&
  {Flock}}]{Chrenko_2022}
{Chrenko}, O., {Chametla}, R.~O., {Nesvorn{\'y}}, D., \& {Flock}, M. 2022,
  \aap, 666, A63, \dodoi{10.1051/0004-6361/202244461}

\bibitem[{{Christ} {et~al.}(2019){Christ}, {Montet}, \&
  {Fabrycky}}]{Christ_2019}
{Christ}, C.~N., {Montet}, B.~T., \& {Fabrycky}, D.~C. 2019, \aj, 157, 235,
  \dodoi{10.3847/1538-3881/ab1aae}

\bibitem[{{Cochran} {et~al.}(2011){Cochran}, {Fabrycky}, {Torres}, {Fressin},
  {D{\'e}sert}, {Ragozzine}, {Sasselov}, {Fortney}, {Rowe}, {Brugamyer},
  {Bryson}, {Carter}, {Ciardi}, {Howell}, {Steffen}, {Borucki}, {Koch}, {Winn},
  {Welsh}, {Uddin}, {Tenenbaum}, {Still}, {Seager}, {Quinn}, {Mullally},
  {Miller}, {Marcy}, {MacQueen}, {Lucas}, {Lissauer}, {Latham}, {Knutson},
  {Kinemuchi}, {Johnson}, {Jenkins}, {Isaacson}, {Howard}, {Horch}, {Holman},
  {Henze}, {Haas}, {Gilliland}, {Gautier}, {Ford}, {Fischer}, {Everett},
  {Endl}, {Demory}, {Deming}, {Charbonneau}, {Caldwell}, {Buchhave}, {Brown},
  \& {Batalha}}]{Cochran_2011}
{Cochran}, W.~D., {Fabrycky}, D.~C., {Torres}, G., {et~al.} 2011, \apjs, 197,
  7, \dodoi{10.1088/0067-0049/197/1/7}

\bibitem[{Dawson(2020)}]{Dawson_2020}
Dawson, R.~I. 2020, \aj, 159, 223, \dodoi{10.3847/1538-3881/ab7fa5}

\bibitem[{{Dawson} {et~al.}(2015){Dawson}, {Murray-Clay}, \&
  {Johnson}}]{Dawson_2015}
{Dawson}, R.~I., {Murray-Clay}, R.~A., \& {Johnson}, J.~A. 2015, \apj, 798, 66,
  \dodoi{10.1088/0004-637X/798/2/66}

\bibitem[{Dawson {et~al.}(2019)Dawson, Huang, Lissauer, Collins, Sha,
  Armstrong, Conti, Collins, Evans, Gan, Horne, Ireland, Murgas, Myers, Relles,
  Sefako, Shporer, Stockdale, {\v{Z}}erjal, Zhou, Ricker, Vanderspek, Latham,
  Seager, Winn, Jenkins, Bouma, Caldwell, Daylan, Doty, Dynes, Esquerdo, Rose,
  Smith, \& Yu}]{Dawson_2019}
Dawson, R.~I., Huang, C.~X., Lissauer, J.~J., {et~al.} 2019, \aj, 158, 65,
  \dodoi{10.3847/1538-3881/ab24ba}

\bibitem[{Dawson {et~al.}(2021)Dawson, Huang, Brahm, Collins, Hobson,
  Jord{\'{a}}n, Dong, Korth, Trifonov, Abe, Agabi, Bruni, Butler, Barbieri,
  Collins, Conti, Crane, Crouzet, Dransfield, Evans, Espinoza, Gan, Guillot,
  Henning, Lissauer, Jensen, Sainte, M{\'{e}}karnia, Myers, Nandakumar, Relles,
  Sarkis, Torres, Shectman, Schmider, Shporer, Stockdale, Teske, Triaud, Wang,
  Ziegler, Ricker, Vanderspek, Latham, Seager, Winn, Jenkins, Bouma, Burt,
  Charbonneau, Levine, McDermott, McLean, Rose, Vanderburg, \&
  Wohler}]{Dawson_2021}
Dawson, R.~I., Huang, C.~X., Brahm, R., {et~al.} 2021, \aj, 161, 161,
  \dodoi{10.3847/1538-3881/abd8d0}

\bibitem[{Deck \& Agol(2015)}]{Deck_2015}
Deck, K.~M., \& Agol, E. 2015, \apj, 802, 116,
  \dodoi{10.1088/0004-637x/802/2/116}

\bibitem[{Deck {et~al.}(2014)Deck, Agol, Holman, \& Nesvorn{\'{y}
  }}]{Deck_2014}
Deck, K.~M., Agol, E., Holman, M.~J., \& Nesvorn{\'{y} }, D. 2014, \apj, 787,
  132, \dodoi{10.1088/0004-637x/787/2/132}

\bibitem[{Deeg \& Alonso(2018)}]{Deeg_2018}
Deeg, H.~J., \& Alonso, R. 2018, in Handbook of Exoplanets (Springer
  International Publishing), 633--657, \dodoi{10.1007/978-3-319-55333-7_117}

\bibitem[{{Fabrycky} \& {Tremaine}(2007)}]{Fabrycky_2007}
{Fabrycky}, D., \& {Tremaine}, S. 2007, \apj, 669, 1298, \dodoi{10.1086/521702}

\bibitem[{{Fausnaugh} {et~al.}(2020){Fausnaugh}, {Burke}, {Ricker}, \&
  {Vanderspek}}]{Fausnaugh_2020}
{Fausnaugh}, M.~M., {Burke}, C.~J., {Ricker}, G.~R., \& {Vanderspek}, R. 2020,
  Research Notes of the American Astronomical Society, 4, 251,
  \dodoi{10.3847/2515-5172/abd63a}

\bibitem[{Foreman-Mackey(2016)}]{corner}
Foreman-Mackey, D. 2016, The Journal of Open Source Software, 1, 24,
  \dodoi{10.21105/joss.00024}

\bibitem[{Foreman-Mackey {et~al.}(2013)Foreman-Mackey, Hogg, Lang, \&
  Goodman}]{Foreman_Mackey_2013}
Foreman-Mackey, D., Hogg, D.~W., Lang, D., \& Goodman, J. 2013, \pasp, 125,
  306, \dodoi{10.1086/670067}

\bibitem[{{Foreman-Mackey} {et~al.}(2021{\natexlab{a}}){Foreman-Mackey},
  {Luger}, {Agol}, {Barclay}, {Bouma}, {Brandt}, {Czekala}, {David}, {Dong},
  {Gilbert}, {Gordon}, {Hedges}, {Hey}, {Morris}, {Price-Whelan}, \&
  {Savel}}]{exoplanet:joss}
{Foreman-Mackey}, D., {Luger}, R., {Agol}, E., {et~al.} 2021{\natexlab{a}},
  arXiv e-prints, arXiv:2105.01994.
\newblock \doarXiv{2105.01994}

\bibitem[{{Foreman-Mackey} {et~al.}(2021{\natexlab{b}}){Foreman-Mackey}, Savel,
  Luger, Agol, Czekala, Price-Whelan, Hedges, Gilbert, Bouma, Brandt, \&
  Barclay}]{exoplanet:zenodo}
{Foreman-Mackey}, D., Savel, A., Luger, R., {et~al.} 2021{\natexlab{b}},
  exoplanet-dev/exoplanet v0.5.1, \dodoi{10.5281/zenodo.1998447}

\bibitem[{Freudenthal {et~al.}(2018)Freudenthal, von Essen, Dreizler,
  Wedemeyer, Agol, Morris, Becker, Mallonn, Hoyer, Ofir, Tal-Or, Deeg, Herrero,
  Ribas, Khalafinejad, Hern{\'{a} }ndez, \& S.}]{Freudenthal_2018}
Freudenthal, J., von Essen, C., Dreizler, S., {et~al.} 2018, \aap, 618, A41,
  \dodoi{10.1051/0004-6361/201833436}

\bibitem[{{Gao} \& {Zhang}(2020)}]{Gao_2020}
{Gao}, P., \& {Zhang}, X. 2020, \apj, 890, 93, \dodoi{10.3847/1538-4357/ab6a9b}

\bibitem[{Gilbert(2022)}]{Gilbert_2022}
Gilbert, G.~J. 2022, \aj, 163, 111, \dodoi{10.3847/1538-3881/ac45f4}

\bibitem[{{Ginsburg} {et~al.}(2019){Ginsburg}, {Sip{\H{o}}cz}, {Brasseur},
  {Cowperthwaite}, {Craig}, {Deil}, {Guillochon}, {Guzman}, {Liedtke}, {Lian
  Lim}, {Lockhart}, {Mommert}, {Morris}, {Norman}, {Parikh}, {Persson},
  {Robitaille}, {Segovia}, {Singer}, {Tollerud}, {de Val-Borro}, {Valtchanov},
  {Woillez}, {Astroquery Collaboration}, \& {a subset of astropy
  Collaboration}}]{astroquery}
{Ginsburg}, A., {Sip{\H{o}}cz}, B.~M., {Brasseur}, C.~E., {et~al.} 2019, \aj,
  157, 98, \dodoi{10.3847/1538-3881/aafc33}

\bibitem[{{Goldberg} {et~al.}(2019){Goldberg}, {Hadden}, {Payne}, \&
  {Holman}}]{Goldberg_2019}
{Goldberg}, M., {Hadden}, S., {Payne}, M.~J., \& {Holman}, M.~J. 2019, \aj,
  157, 142, \dodoi{10.3847/1538-3881/ab06c9}

\bibitem[{{Goldreich} \& {Tremaine}(1980)}]{Goldreich_1980}
{Goldreich}, P., \& {Tremaine}, S. 1980, \apj, 241, 425, \dodoi{10.1086/158356}

\bibitem[{{Goodman} \& {Weare}(2010)}]{MCMC}
{Goodman}, J., \& {Weare}, J. 2010, Communications in Applied Mathematics and
  Computational Science, 5, 65, \dodoi{10.2140/camcos.2010.5.65}

\bibitem[{{Hadden} \& {Lithwick}(2017)}]{Hadden_2017}
{Hadden}, S., \& {Lithwick}, Y. 2017, \aj, 154, 5,
  \dodoi{10.3847/1538-3881/aa71ef}

\bibitem[{Hamann {et~al.}(2019)Hamann, Montet, Fabrycky, Agol, \&
  Kruse}]{Hamann_2019}
Hamann, A., Montet, B.~T., Fabrycky, D.~C., Agol, E., \& Kruse, E. 2019, \aj,
  158, 133, \dodoi{10.3847/1538-3881/ab32e3}

\bibitem[{Harris {et~al.}(2020)Harris, Millman, van~der Walt, Gommers,
  Virtanen, Cournapeau, Wieser, Taylor, Berg, Smith, Kern, Picus, Hoyer, van
  Kerkwijk, Brett, Haldane, del R{\'{i}}o, Wiebe, Peterson,
  G{\'{e}}rard-Marchant, Sheppard, Reddy, Weckesser, Abbasi, Gohlke, \&
  Oliphant}]{numpy}
Harris, C.~R., Millman, K.~J., van~der Walt, S.~J., {et~al.} 2020, Nature, 585,
  357, \dodoi{10.1038/s41586-020-2649-2}

\bibitem[{{Hirano} {et~al.}(2018){Hirano}, {Dai}, {Gandolfi}, {Fukui},
  {Livingston}, {Miyakawa}, {Endl}, {Cochran}, {Alonso-Floriano}, {Kuzuhara},
  {Montes}, {Ryu}, {Albrecht}, {Barragan}, {Cabrera}, {Csizmadia}, {Deeg},
  {Eigm{\"u}ller}, {Erikson}, {Fridlund}, {Grziwa}, {Guenther}, {Hatzes},
  {Korth}, {Kudo}, {Kusakabe}, {Narita}, {Nespral}, {Nowak}, {P{\"a}tzold},
  {Palle}, {Persson}, {Prieto-Arranz}, {Rauer}, {Ribas}, {Sato}, {Smith},
  {Tamura}, {Tanaka}, {Van Eylen}, \& {Winn}}]{Hirano_2018}
{Hirano}, T., {Dai}, F., {Gandolfi}, D., {et~al.} 2018, \aj, 155, 127,
  \dodoi{10.3847/1538-3881/aaa9c1}

\bibitem[{{Holman} {et~al.}(2010){Holman}, {Fabrycky}, {Ragozzine}, {Ford},
  {Steffen}, {Welsh}, {Lissauer}, {Latham}, {Marcy}, {Walkowicz}, {Batalha},
  {Jenkins}, {Rowe}, {Cochran}, {Fressin}, {Torres}, {Buchhave}, {Sasselov},
  {Borucki}, {Koch}, {Basri}, {Brown}, {Caldwell}, {Charbonneau}, {Dunham},
  {Gautier}, {Geary}, {Gilliland}, {Haas}, {Howell}, {Ciardi}, {Endl},
  {Fischer}, {F{\"u}r{\'e}sz}, {Hartman}, {Isaacson}, {Johnson}, {MacQueen},
  {Moorhead}, {Morehead}, \& {Orosz}}]{Holman_2010}
{Holman}, M.~J., {Fabrycky}, D.~C., {Ragozzine}, D., {et~al.} 2010, Science,
  330, 51, \dodoi{10.1126/science.1195778}

\bibitem[{{Huang} {et~al.}(2016){Huang}, {Wu}, \& {Triaud}}]{Huang_2016}
{Huang}, C., {Wu}, Y., \& {Triaud}, A. H.~M.~J. 2016, \apj, 825, 98,
  \dodoi{10.3847/0004-637X/825/2/98}

\bibitem[{Hunter(2007)}]{matplotlib}
Hunter, J.~D. 2007, Computing in Science and Engineering, 9, 90,
  \dodoi{10.1109/MCSE.2007.55}

\bibitem[{{Jenkins} {et~al.}(2016){Jenkins}, {Twicken}, {McCauliff},
  {Campbell}, {Sanderfer}, {Lung}, {Mansouri-Samani}, {Girouard}, {Tenenbaum},
  {Klaus}, {Smith}, {Caldwell}, {Chacon}, {Henze}, {Heiges}, {Latham},
  {Morgan}, {Swade}, {Rinehart}, \& {Vanderspek}}]{Jenkins_2016}
{Jenkins}, J.~M., {Twicken}, J.~D., {McCauliff}, S., {et~al.} 2016, in Society
  of Photo-Optical Instrumentation Engineers (SPIE) Conference Series, Vol.
  9913, Software and Cyberinfrastructure for Astronomy IV, ed. G.~{Chiozzi} \&
  J.~C. {Guzman}, 99133E, \dodoi{10.1117/12.2233418}

\bibitem[{{Jontof-Hutter} {et~al.}(2014){Jontof-Hutter}, {Lissauer}, {Rowe}, \&
  {Fabrycky}}]{Jontof-Hutter_2014}
{Jontof-Hutter}, D., {Lissauer}, J.~J., {Rowe}, J.~F., \& {Fabrycky}, D.~C.
  2014, \apj, 785, 15, \dodoi{10.1088/0004-637X/785/1/15}

\bibitem[{Kipping {et~al.}(2019)Kipping, Nesvorn{\'{y} }, Hartman, Torres,
  Bakos, Jansen, \& Teachey}]{Kipping_2019}
Kipping, D., Nesvorn{\'{y} }, D., Hartman, J., {et~al.} 2019, \mnras, 486,
  4980, \dodoi{10.1093/mnras/stz1141}

\bibitem[{Kipping(2013)}]{Kipping_2013}
Kipping, D.~M. 2013, \mnras, 435, 2152, \dodoi{10.1093/mnras/stt1435}

\bibitem[{Kumar {et~al.}(2019)Kumar, Carroll, Hartikainen, \&
  Martin}]{exoplanet:arviz}
Kumar, R., Carroll, C., Hartikainen, A., \& Martin, O.~A. 2019, The Journal of
  Open Source Software, \dodoi{10.21105/joss.01143}

\bibitem[{Lee \& Chiang(2016)}]{Lee_2016}
Lee, E.~J., \& Chiang, E. 2016, \apj, 817, 90,
  \dodoi{10.3847/0004-637x/817/2/90}

\bibitem[{{Liang} {et~al.}(2021){Liang}, {Robnik}, \& {Seljak}}]{Liang_2021}
{Liang}, Y., {Robnik}, J., \& {Seljak}, U. 2021, \aj, 161, 202,
  \dodoi{10.3847/1538-3881/abe6a7}

\bibitem[{{Lightkurve Collaboration} {et~al.}(2018){Lightkurve Collaboration},
  {Cardoso}, {Hedges}, {Gully-Santiago}, {Saunders}, {Cody}, {Barclay}, {Hall},
  {Sagear}, {Turtelboom}, {Zhang}, {Tzanidakis}, {Mighell}, {Coughlin}, {Bell},
  {Berta-Thompson}, {Williams}, {Dotson}, \& {Barentsen}}]{lightkurve}
{Lightkurve Collaboration}, {Cardoso}, J.~V.~d.~M., {Hedges}, C., {et~al.}
  2018, {Lightkurve: Kepler and TESS time series analysis in Python},
  Astrophysics Source Code Library.
\newblock \doeprint{1812.013}

\bibitem[{{Lin} \& {Papaloizou}(1986)}]{Lin_1986}
{Lin}, D.~N.~C., \& {Papaloizou}, J. 1986, \apj, 309, 846,
  \dodoi{10.1086/164653}

\bibitem[{{Lissauer} {et~al.}(2011){Lissauer}, {Fabrycky}, {Ford}, {Borucki},
  {Fressin}, {Marcy}, {Orosz}, {Rowe}, {Torres}, {Welsh}, {Batalha}, {Bryson},
  {Buchhave}, {Caldwell}, {Carter}, {Charbonneau}, {Christiansen}, {Cochran},
  {Desert}, {Dunham}, {Fanelli}, {Fortney}, {Gautier}, {Geary}, {Gilliland},
  {Haas}, {Hall}, {Holman}, {Koch}, {Latham}, {Lopez}, {McCauliff}, {Miller},
  {Morehead}, {Quintana}, {Ragozzine}, {Sasselov}, {Short}, \&
  {Steffen}}]{Lissauer_2011}
{Lissauer}, J.~J., {Fabrycky}, D.~C., {Ford}, E.~B., {et~al.} 2011, \nat, 470,
  53, \dodoi{10.1038/nature09760}

\bibitem[{Lithwick {et~al.}(2012)Lithwick, Xie, \& Wu}]{Lithwick_2012}
Lithwick, Y., Xie, J., \& Wu, Y. 2012, \apj, 761, 122,
  \dodoi{10.1088/0004-637x/761/2/122}

\bibitem[{{Lopez} \& {Fortney}(2014)}]{Lopez_2014}
{Lopez}, E.~D., \& {Fortney}, J.~J. 2014, \apj, 792, 1,
  \dodoi{10.1088/0004-637X/792/1/1}

\bibitem[{{Luger} {et~al.}(2019){Luger}, {Agol}, {Foreman-Mackey}, {Fleming},
  {Lustig-Yaeger}, \& {Deitrick}}]{exoplanet:luger18}
{Luger}, R., {Agol}, E., {Foreman-Mackey}, D., {et~al.} 2019, \aj, 157, 64,
  \dodoi{10.3847/1538-3881/aae8e5}

\bibitem[{{Masuda}(2017)}]{Masuda_2017}
{Masuda}, K. 2017, \aj, 154, 64, \dodoi{10.3847/1538-3881/aa7aeb}

\bibitem[{{Mills} \& {Fabrycky}(2017)}]{Mills_Fabrycky_2017}
{Mills}, S.~M., \& {Fabrycky}, D.~C. 2017, \aj, 153, 45,
  \dodoi{10.3847/1538-3881/153/1/45}

\bibitem[{{Mills} \& {Mazeh}(2017)}]{Mills_Mazeh_2017}
{Mills}, S.~M., \& {Mazeh}, T. 2017, \apjl, 839, L8,
  \dodoi{10.3847/2041-8213/aa67eb}

\bibitem[{Miralda-Escude(2002)}]{Miralda_Escude_2002}
Miralda-Escude, J. 2002, \apj, 564, 1019, \dodoi{10.1086/324279}

\bibitem[{{Nesvorn{\'y}} {et~al.}(2022){Nesvorn{\'y}}, {Chrenko}, \&
  {Flock}}]{Nesvorny_2022}
{Nesvorn{\'y}}, D., {Chrenko}, O., \& {Flock}, M. 2022, \apj, 925, 38,
  \dodoi{10.3847/1538-4357/ac36cd}

\bibitem[{{Nesvorn{\'y}} {et~al.}(2013){Nesvorn{\'y}}, {Kipping}, {Terrell},
  {Hartman}, {Bakos}, \& {Buchhave}}]{Nesvorny_2013}
{Nesvorn{\'y}}, D., {Kipping}, D., {Terrell}, D., {et~al.} 2013, \apj, 777, 3,
  \dodoi{10.1088/0004-637X/777/1/3}

\bibitem[{Nesvorn{\'{y} } \& Vokrouhlick{\'{y}}(2016)}]{Nesvorny_2016}
Nesvorn{\'{y} }, D., \& Vokrouhlick{\'{y}}, D. 2016, \apj, 823, 72,
  \dodoi{10.3847/0004-637x/823/2/72}

\bibitem[{{Ohno} \& {Tanaka}(2021)}]{Ohno_2021}
{Ohno}, K., \& {Tanaka}, Y.~A. 2021, \apj, 920, 124,
  \dodoi{10.3847/1538-4357/ac1516}

\bibitem[{{P{\'a}l}(2008)}]{Pal_2008}
{P{\'a}l}, A. 2008, \mnras, 390, 281, \dodoi{10.1111/j.1365-2966.2008.13723.x}

\bibitem[{{Piaulet} {et~al.}(2021){Piaulet}, {Benneke}, {Rubenzahl}, {Howard},
  {Lee}, {Thorngren}, {Angus}, {Peterson}, {Schlieder}, {Werner}, {Kreidberg},
  {Jaouni}, {Crossfield}, {Ciardi}, {Petigura}, {Livingston}, {Dressing},
  {Fulton}, {Beichman}, {Christiansen}, {Gorjian}, {Hardegree-Ullman}, {Krick},
  \& {Sinukoff}}]{Piaulet_2021}
{Piaulet}, C., {Benneke}, B., {Rubenzahl}, R.~A., {et~al.} 2021, \aj, 161, 70,
  \dodoi{10.3847/1538-3881/abcd3c}

\bibitem[{{Piro} \& {Vissapragada}(2020)}]{Piro_2020}
{Piro}, A.~L., \& {Vissapragada}, S. 2020, \aj, 159, 131,
  \dodoi{10.3847/1538-3881/ab7192}

\bibitem[{{Rein} \& {Liu}(2012)}]{rebound}
{Rein}, H., \& {Liu}, S.~F. 2012, \aap, 537, A128,
  \dodoi{10.1051/0004-6361/201118085}

\bibitem[{{Rein} \& {Spiegel}(2015)}]{reboundias15}
{Rein}, H., \& {Spiegel}, D.~S. 2015, \mnras, 446, 1424,
  \dodoi{10.1093/mnras/stu2164}

\bibitem[{{Ricker} {et~al.}(2015){Ricker}, {Winn}, {Vanderspek}, {Latham},
  {Bakos}, {Bean}, {Berta-Thompson}, {Brown}, {Buchhave}, {Butler}, {Butler},
  {Chaplin}, {Charbonneau}, {Christensen-Dalsgaard}, {Clampin}, {Deming},
  {Doty}, {De Lee}, {Dressing}, {Dunham}, {Endl}, {Fressin}, {Ge}, {Henning},
  {Holman}, {Howard}, {Ida}, {Jenkins}, {Jernigan}, {Johnson}, {Kaltenegger},
  {Kawai}, {Kjeldsen}, {Laughlin}, {Levine}, {Lin}, {Lissauer}, {MacQueen},
  {Marcy}, {McCullough}, {Morton}, {Narita}, {Paegert}, {Palle}, {Pepe},
  {Pepper}, {Quirrenbach}, {Rinehart}, {Sasselov}, {Sato}, {Seager},
  {Sozzetti}, {Stassun}, {Sullivan}, {Szentgyorgyi}, {Torres}, {Udry}, \&
  {Villasenor}}]{Ricker_2015}
{Ricker}, G.~R., {Winn}, J.~N., {Vanderspek}, R., {et~al.} 2015, Journal of
  Astronomical Telescopes, Instruments, and Systems, 1, 014003,
  \dodoi{10.1117/1.JATIS.1.1.014003}

\bibitem[{Salvatier {et~al.}(2016)Salvatier, Wiecki, \&
  Fonnesbeck}]{exoplanet:pymc3}
Salvatier, J., Wiecki, T.~V., \& Fonnesbeck, C. 2016, PeerJ Computer Science,
  2, e55

\bibitem[{Shahaf {et~al.}(2021)Shahaf, Mazeh, Zucker, \&
  Fabrycky}]{Shahaf_2021}
Shahaf, S., Mazeh, T., Zucker, S., \& Fabrycky, D. 2021, \mnras, 505, 1293,
  \dodoi{10.1093/mnras/stab1359}

\bibitem[{Smith {et~al.}(2012)Smith, Stumpe, Cleve, Jenkins, Barclay, Fanelli,
  Girouard, Kolodziejczak, McCauliff, Morris, \& Twicken}]{Smith_2012}
Smith, J.~C., Stumpe, M.~C., Cleve, J. E.~V., {et~al.} 2012, \pasp, 124, 1000,
  \dodoi{10.1086/667697}

\bibitem[{{Steffen}(2016)}]{Steffen_2016}
{Steffen}, J.~H. 2016, \mnras, 457, 4384, \dodoi{10.1093/mnras/stw241}

\bibitem[{{Stumpe} {et~al.}(2014){Stumpe}, {Smith}, {Catanzarite}, {Van Cleve},
  {Jenkins}, {Twicken}, \& {Girouard}}]{Stumpe_2014}
{Stumpe}, M.~C., {Smith}, J.~C., {Catanzarite}, J.~H., {et~al.} 2014, \pasp,
  126, 100, \dodoi{10.1086/674989}

\bibitem[{{Theano Development Team}(2016)}]{exoplanet:theano}
{Theano Development Team}. 2016, arXiv e-prints, abs/1605.02688.
\newblock \url{http://arxiv.org/abs/1605.02688}

\bibitem[{Virtanen {et~al.}(2020)Virtanen, Gommers, Oliphant, Haberland, Reddy,
  Cournapeau, Burovski, Peterson, Weckesser, Bright, {van der Walt}, Brett,
  Wilson, Millman, Mayorov, Nelson, Jones, Kern, Larson, Carey, Polat, Feng,
  Moore, {VanderPlas}, Laxalde, Perktold, Cimrman, Henriksen, Quintero, Harris,
  Archibald, Ribeiro, Pedregosa, {van Mulbregt}, \& {SciPy 1.0
  Contributors}}]{scipy}
Virtanen, P., Gommers, R., Oliphant, T.~E., {et~al.} 2020, Nature Methods, 17,
  261, \dodoi{10.1038/s41592-019-0686-2}

\bibitem[{{Weiss} \& {Marcy}(2014)}]{Weiss_2014}
{Weiss}, L.~M., \& {Marcy}, G.~W. 2014, \apjl, 783, L6,
  \dodoi{10.1088/2041-8205/783/1/L6}

\bibitem[{{Winn}(2010)}]{Winn_2010}
{Winn}, J.~N. 2010, arXiv e-prints, arXiv:1001.2010.
\newblock \doarXiv{1001.2010}

\end{thebibliography}
\bibliographystyle{aasjournal}



\end{document}